\documentclass[fleqn,usenatbib]{mnras}

\usepackage{newtxtext,newtxmath}
\usepackage[T1]{fontenc}

\DeclareRobustCommand{\VAN}[3]{#2}
\let\VANthebibliography\thebibliography
\def\thebibliography{\DeclareRobustCommand{\VAN}[3]{##3}\VANthebibliography}

\usepackage{graphicx}	% Including figure files
\usepackage{amsmath}
\usepackage{CJKutf8}
\usepackage{comment}
\usepackage{fix-cm} 
\usepackage[normalem]{ulem}
\newcommand{\YWchinesename}{{\begin{CJK}{UTF8}{gbsn}(王颖翔)\end{CJK}}}
\newcommand{\YLchinesename}{{\begin{CJK}{UTF8}{gbsn}(李亚光)\end{CJK}}}
\newcommand{\YCchinesename}{{\begin{CJK}{UTF8}{gbsn}(陈逸凡)\end{CJK}}}
\renewcommand{\labelenumi}{(\arabic{enumi})}

\newcommand{\kepler}[0]{\emph{Kepler}}
\newcommand{\corot}[0]{\emph{CoRoT}}
\newcommand{\tess}[0]{\emph{TESS}}

\newcommand{\Dnu}[0]{$\Delta\nu$}

\newcommand{\numax}[0]{\mbox{$\nu_{\rm max}$}}
\newcommand{\eps}[0]{$\varepsilon$}
\newcommand{\dnuoz}[0]{$\delta\nu_{02}$}
\newcommand{\dnuol}[0]{$\delta\nu_{01}$}

\title[Phase Shifts \& Small Separations in Kepler Giants]{Asteroseismic Diagnostics for Red Giants with Kepler: Measuring \eps{} and Small Frequency Separations in 16,000 Stars}

\author[Y. Wang et al.]{
Yingxiang Wang\YWchinesename,$^{1}$\thanks{E-mail: ywan0509@uni.sydney.edu.au}
Timothy R. Bedding,$^{1}$
Yaguang Li\YLchinesename,$^{2}$
Yifan Chen\YCchinesename,$^{1}$
\newauthor
Courtney L. Crawford,$^{1}$
Daniel Huber$^{2}$ and
K. R. Sreenivas$^{1}$
\\
$^{1}$Sydney Institute for Astronomy (SIfA), School of Physics, University of Sydney, NSW 2006, Australia\\
$^{2}$Institute for Astronomy, University of Hawai`i, 2680 Woodlawn Drive, Honolulu, HI 96822, USA
}

\date{Accepted XXX. Received YYY; in original form ZZZ}

\pubyear{\the\year{}}

\begin{document}
\label{firstpage}
\pagerange{\pageref{firstpage}--\pageref{lastpage}}
\maketitle

\begin{abstract}
Asteroseismic studies of red giants have primarily relied on two global parameters: the large frequency separation (\(\Delta\nu\)) and the frequency of maximum power (\(\nu_{\rm max}\)). Meanwhile, the p-mode phase shift (\(\varepsilon\)) and small frequency separations (\(\delta\nu_{01}\), \(\delta\nu_{02}\)), which offer additional constraints on stellar interiors, remain underexplored due to measurement challenges. Here we develop an automated pipeline based on collapsed \'echelle diagrams and apply it to \(\sim 16{,}000\) \emph{Kepler} red giants, jointly measuring \(\Delta\nu\), \(\varepsilon\), \(\delta\nu_{01}\), and \(\delta\nu_{02}\) and assembling the largest homogeneous catalogue of these quantities to date, together with updated \(\Delta\nu\) values and formal internal uncertainties. Using this catalogue, we quantify evolutionary trends across the red-giant branch and core-helium-burning phase. We find that \(\delta\nu_{02}/\Delta\nu\) stays nearly constant for RGB stars and, for core-helium-burning stars, organises into two sequences that are systematically offset but partially overlap, broadly separating stars in the red-clump and secondary-clump regimes. We also trace the mass- and metallicity-dependent helium-flash transition. Meanwhile, \(\varepsilon\) follows a single \(\Delta\nu\)–\(\varepsilon\) relation common to both evolutionary phases. Comparisons with stellar-evolution models reveal systematic offsets in \(\varepsilon\) and \(\delta\nu_{01}\), which we interpret as signatures of near-surface and outer-envelope modelling deficiencies. These comparisons further suggest that dipole-mode small separations are sensitive to mode-dependent surface terms in evolved stars. Overall, our results demonstrate that \(\varepsilon\) and the small separations provide important diagnostics of core structure, convective-boundary mixing, and helium ignition that are complementary to those provided by \(\Delta\nu\) and \(\nu_{\rm max}\) alone. The resulting catalogue offers a reference for testing and calibrating future stellar-evolution models.

\end{abstract}


%%%%%%%%%%%%%%%%%%%%%%%%%%%%%%%%%%%%%%%%%%%%%%%%%%

%%%%%%%%%%%%%%%%% BODY OF PAPER %%%%%%%%%%%%%%%%%%

\section{Introduction}

Asteroseismic analysis of red giants provides critical insights into stellar evolution through the detection of oscillations. Space-based photometric missions such as \corot{} \citep{Baglin2006cosp...36.3749B}, \kepler{} \citep{Borucki2008Kepler}, and \tess{} \citep{Ricker2015JATIS...1a4003R} have enabled systematic studies of these oscillations (see reviews by \citealt{Chaplin2013ARA&A..51..353C,Hekker2017A&ARv..25....1H,Jackiewicz2021FrASS...7..102J}). While asteroseismic techniques can, in  principle, use full sets of stellar oscillation frequencies, population-scale analyses often rely on two key global parameters: the large frequency separation (\Dnu), and the frequency of maximum oscillation power (\numax). These parameters have yielded precise mass and radius determinations for over 16,000 red giants in \kepler{} Data Release 25 (DR25) \citep{Mosser2010A&A...517A..22M,Yu2018,Hon2024ApJ...973..154H,Pinsonneault2025}.

However, beyond the two widely used global parameters \Dnu\ and \numax, additional seismic parameters such as the small frequency separations (\(\delta\nu_{0l}\)) and the phase shift (\(\varepsilon\)) have received comparatively less attention in large samples of red giants. These parameters are defined by the asymptotic relation for acoustic modes \citep{Tassoul1980ApJS...43..469T}:
\begin{equation}
    \nu_{n,l} \approx \Delta\nu\left(n+\frac{l}{2}+\varepsilon\right)-\delta\nu_{0 l},
    \label{equa:asymptotic}
\end{equation}
where \(n\) denotes the radial order, \(l\) is the spherical degree, and \(\delta\nu_{0 l}\) represents spacings between non-radial (\(l\neq0\)) and radial (\(l=0\)) p-modes. In observations, there are two small separations that can be estimated in a large number of solar-like oscillators: \(\delta\nu_{02}\) and \(\delta\nu_{01}\). To be explicit, for a given radial order, \(n\), these separations are defined as follows:
\begin{subequations}
    \begin{align}
        \delta\nu_{02}&=\nu_{n,l=0}-\nu_{n-1,l=2} \\
        \delta\nu_{01}&=\frac{1}{2}(\nu_{n,l=0}+\nu_{n+1,l=0})-\nu_{n,l=1}
    \end{align}
\end{subequations}
For main-sequence stars, these parameters are established as direct diagnostics of interior structure, and depend on masses and ages of solar-like oscillators (e.g. \citealt{Roxburgh2003A&A...411..215R,Roxburgh2005A&A...434..665R,Floranes2005MNRAS.356..671O}). Such an interpretation is based on a well-known integral estimator relating \dnuoz{} to a radially averaged sound-speed gradient. However, for red giants, the small separations are much less sensitive to their internal structure \citep{Montalban2010ApJL721L182, Christensen-Dalsgaard2014aste.book..194C}.  

Recently, \eps{} and \dnuoz{} have been reported to help constrain the properties of convective boundary mixing, especially when undershooting occurs below the convective envelope \citep{Ong2025ApJ...980..199O,Reyes2025Natur.640..338R}. However, for red giants, systematic studies of \dnuol{} and \dnuoz{} still remain confined to small samples \citep{Huber2010,Corsaro2012ApJ...757..190C}.\footnote{Values of \dnuoz{} were measured for about 6200 Kepler red giants by \citet{Kallinger2019arXiv190609428K} but no analysis has been published.} This reveals potential to expand high-precision \dnuol{}, \dnuoz{} and \eps{} measurements to larger samples, strengthening connections between stellar models and observations. Complementary \emph{TESS} results with individual frequencies and asymptotic parameters for red giants have recently been presented by \citet{Zhou2025ApJS27937}, providing an ensemble view that complements our Kepler-based catalogue.

This work aims to extend high-precision determinations of \dnuol{}, \dnuoz{} and \eps{} to all \(\sim 16{,}000\) \kepler{} red giants, and to refine \Dnu{} for the same sample, thereby improving mass estimates via the seismic scaling relations. By analysing these parameters across evolutionary phases, we investigate their correlations with fundamental stellar properties and test theoretical models linking seismic observables to internal structure. Our results demonstrate that small separations and phase shifts offer critical constraints on stellar evolution not accessible through \Dnu{} and \numax{} alone. Using collapsed \'echelle diagrams, we measure phase shifts (\eps) and small frequency separations (\dnuol, \dnuoz) for \(\sim 16{,}000\) \kepler{} red giants in Section~\ref{sec:Method}. At the same time, we also produce refined measurements of \(\Delta\nu\). Section~\ref{sec:result} presents our seismic catalogue. In Section~\ref{sec:Mass}, we compare observations with models. Finally, Section~\ref{sec:Conclusions} summarizes our findings and their implications.

\section{Methodology}\label{sec:Method}

\begin{figure*}
    \centering
    \includegraphics[width=\textwidth]{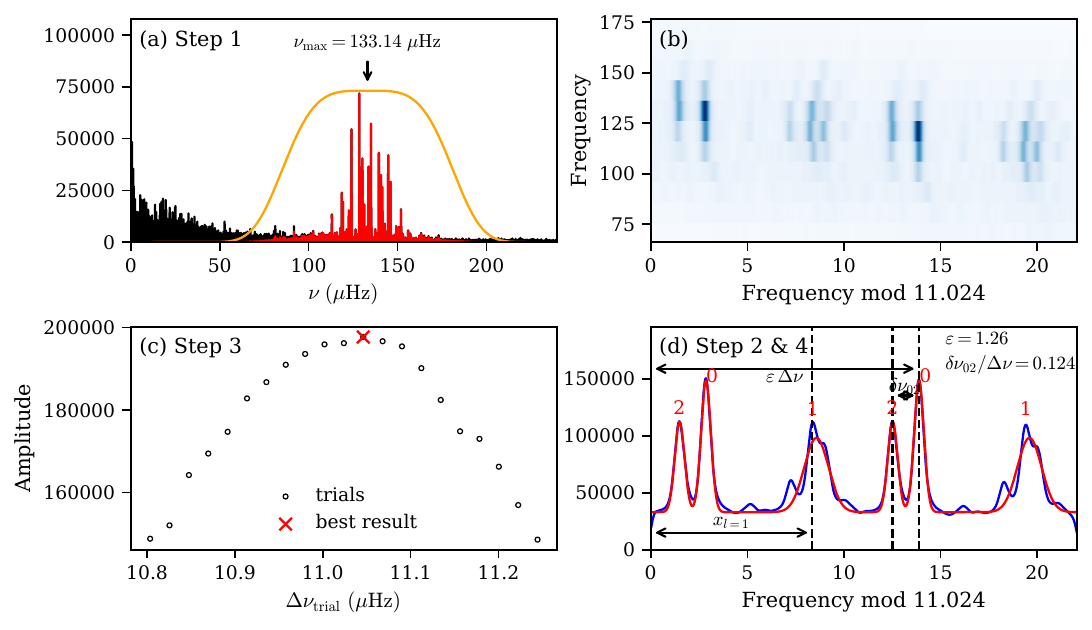}
    \caption{\protect\unboldmath Illustration of our collapsed-\'{e}chelle method for a representative \textit{Kepler} red giant (KIC 3530333). 
    (a) Pre-processed PDS (black), super-Gaussian window function (yellow), and the filtered spectrum used for the analysis (red). 
    (b) \'{E}chelle diagram constructed from the filtered spectrum using the adopted large frequency separation $\Delta\nu$, shown over a frequency range of $2\Delta\nu$ to reduce edge effects. 
    (c) Metric (see Eq.~\ref{metric}) as a function of trial $\Delta\nu$ values; open circles show the metric for each trial, and the red cross marks the adopted value corresponding to the maximum metric. 
    (d) Collapsed \'{e}chelle diagram for the adopted $\Delta\nu$. Each peak corresponds to a mode ridge in panel (b), and the symbols indicate the quantities measured from this diagram ($\epsilon$ and the small separations).}
    \label{fig:Collapsed}
\end{figure*}

The method we used to estimate \(\Delta\nu\), \(\varepsilon\), \(\delta\nu_{02}\), and \(\delta\nu_{01}\) can be summarized in the following steps (see also Fig.~\ref{fig:Collapsed}):
{
\renewcommand{\labelenumi}{(\arabic{enumi})}
\begin{enumerate}
    \item \textbf{Prepare the oscillation power-density spectrum}: Remove background noise and adjust extreme peaks to ensure balanced mode amplitudes (Section~\ref{step:prep_spectrum}).
    
    \item \textbf{Create the collapsed \'echelle diagram}: Fold the processed spectrum at \(\Delta\nu\) intervals (Section~\ref{subsec:collapsed_echelle}).
    
    \item \textbf{Optimise the large frequency separation}: Sample trial values of \(\Delta\nu\) and select the one that maximises ridge prominence in the collapsed diagram, ensuring consistent global structure (Section~\ref{subsec:dnu_estimation}).
    
    \item \textbf{Extract the seismic parameters}: Perform constrained fitting of ridge centroids and widths to measure \(\varepsilon\), \(\delta\nu_{02}\), and \(\delta\nu_{01}\) from the optimised collapsed diagram (Section~\ref{subsec:final_params}).
\end{enumerate}
}

\subsection{Preparation of Oscillation Power-Density Spectrum}\label{step:prep_spectrum}

To prepare the power-density spectrum (PDS) for analysis, we started from PDS computed from \kepler{} long-cadence PDCSAP light curves processed with the nuSYD pipeline \citep{Sreenivas2024}. No additional time-domain detrending beyond that pipeline was applied at this stage. We then made two adjustments to the raw PDS:

\subsubsection{Isolating the region of oscillation power}\label{Noise Suppression}
Noise outside the oscillation region biases the collapsed \'echelle diagram: after folding the PDS modulo \Dnu{} and averaging across orders, out-of-envelope power elevates the baseline and can introduce spurious local maxima, which pull the centroid of the collapsed profile and hence bias \eps{} (and inflate its uncertainty). To mitigate this, we multiply the PDS by an order-4 super-Gaussian window centred at \numax{}:
\begin{equation}
W(\nu) = \exp\!\left(-\left[\frac{\nu - \numax}{\left(\frac{3\times 0.66}{2.355}\right)\,\numax^{0.88}}\right]^4\right).
\end{equation}
An example is shown in Figure~\ref{fig:Collapsed}(a). Here the window width is tied to the oscillation-power envelope in red giants, which is well described by a Gaussian with full width at half maximum (FWHM) \(\delta\nu_{\rm env} = 0.66\,\numax^{0.88}\) \citep{Mosser2012A&A...537A..30M}. We convert the FWHM to a Gaussian dispersion via \(\sigma \equiv \delta\nu_{\rm env}/2.355\), and adopt a half-width of \(3\sigma\) in the super-Gaussian so that power within the envelope is preserved while out-of-envelope background is sharply attenuated. As the figure illustrates, the super-Gaussian leaves the power near \numax{} essentially unchanged but rapidly suppresses distant frequencies, stabilizing the collapsed profile and the \eps{} estimate.

\subsubsection{Peak Adjustment}  
In red giants, convection stochastically excites and damps oscillation modes \citep{Houdek1999,Samadi_Goupil_2001}, causing amplitude variations. Because of this, some modes occasionally show unusually strong peaks, which can distort measurements in collapsed échelle diagrams (especially \eps{} values) because they disproportionately affect averaged positions. To prevent this distortion, we identified peaks with amplitudes significantly higher than other modes using a median absolute deviation (MAD) rule: a peak of height $h_i$ was flagged when $h_i > \tilde{h} + \kappa\,\mathrm{MAD}$, where $\tilde{h}$ is the sample median across detected peaks and MAD is the median absolute deviation; we used $\kappa=5$. For each unusual peak, we reduced its amplitude to match the strongest neighbouring peak within a symmetric window whose half-width equals \(c\) times the peak’s FWHM, leaving the frequency unchanged.

\subsection{Collapsed \'Echelle Diagram}\label{subsec:collapsed_echelle}

From the prepared oscillation spectrum, we extract global asteroseismic parameters using a collapsed \'echelle diagram. Conceptually, this extends traditional \'echelle analysis \citep{Grec1983} by folding the PDS modulo the large frequency separation, \Dnu, and summing vertically to form a one-dimensional collapsed profile (see Fig.~\ref{fig:Collapsed}b,d). The collapsed idea was already introduced for solar data by \citet{Grec1983} and for solar-like stars \citep[e.g.][]{Bouchy2002A&A...390..205B,Bedding2004ApJ...614..380B}, and a related folded-spectrum method has been used in automated pipelines; see \citet{Zinn2019} for a K2 application. Before folding, we applied two pre-processing steps to improve the visibility of the ridge pattern while preserving frequency separations. As in Section~\ref{Noise Suppression}, we applied the same super-Gaussian window centred on \numax{} to down-weight power far from the oscillation hump. This suppresses granulation/shot-noise outside the envelope. Second, we performed a Gaussian smoothing with a FWHM of \(0.03\,\Delta\nu\) to reduce pixel-to-pixel variance without biasing the relative ridge phases.

The folding process combines power from modes separated by integer multiples of \Dnu. In the collapsed \'echelle, each angular-degree ridge (\(l=0,1,2\)) therefore appears as a \emph{phase-aligned aggregate peak} at its characteristic phase. These peaks correspond to ridge structures in standard \'echelle diagrams [Fig.~\ref{fig:Collapsed}(b)] but reduced to one dimension [Fig.~\ref{fig:Collapsed}(d)]. By focusing on the overall ridge pattern rather than individual mode peaks, the collapsed representation enables robust mode identification and efficient measurement of \dnuoz{}, \dnuol{}, and \eps{} via spectrum stacking and constrained least-squares fitting (see Section~\ref{subsec:dnu_estimation} for the trial-\Dnu\ scan and scoring). To mitigate edge effects, we duplicate the \'echelle pattern by one radial order prior to collapsing \citep{Bedding2012ASPC..462..195B}.

\subsection{\texorpdfstring{\unboldmath Refining the Large Frequency Separation ($\Delta\nu$)}{Refining the Large Frequency Separation (Delta nu)}}\label{subsec:dnu_estimation}

An accurate initial estimate of the large frequency separation, \(\Delta\nu_{0}\), is essential because it sets the folding interval of the collapsed \'echelle diagram and directly affects the reliability of all subsequent parameter measurements. For each star we adopted an initial \(\Delta\nu_{0}\) from the input catalogue and then refined \(\Delta\nu\) by systematically sampling trial values within \(\pm 8\%\) of this value, i.e. from \(0.92\Delta\nu_{0}\) to \(1.08\Delta\nu_{0}\), to allow for potential systematic deviations.

For each trial \(\Delta\nu_{\mathrm{trial}}\), we constructed the collapsed \'echelle diagram using the pre-processed spectrum described in Section~\ref{subsec:collapsed_echelle} and fitted a three-ridge parametric model (\(l=0,1,2\)) plus a constant background while keeping \(\Delta\nu_{\mathrm{trial}}\) fixed. The model profile is
\begin{equation}
M(x)=B+\sum_{l=0}^{2} A_l \sum_{k=-1}^{1}
\exp\!\left[-4\ln 2\,\frac{(x-x_l-k\Delta\nu_{\rm trial})^2}{\Gamma_l^2}\right],
\end{equation}
where \(x_l\) denotes the centroid of the \(l\)-th ridge in the collapsed \'echelle profile, \(A_l\) and \(\Gamma_l\) are the corresponding amplitude and FWHM, and \(B\) is a constant background level. The terms with \(k=\pm1\) are mirror copies included to mitigate edge effects in the collapsed diagram. Parameters were estimated by minimising square-root–weighted residuals between model and data, \((\mathrm{model}-\mathrm{data})/\sqrt{P_{\mathrm{coll}}+10^{-6}}\), where \(P_{\mathrm{coll}}\) denotes the power in the collapsed-\hbox{\'e}chelle spectrum at each pixel (see Section~\ref{subsec:collapsed_echelle}), using the derivative-free Nelder--Mead algorithm as implemented in \texttt{lmfit} \citep{newville_2025_16175987}.

As a scalar measure of ridge prominence, we adopted the metric,
\begin{equation}\label{metric}
\mathcal{R}\equiv \tilde{A}_{0}+\tilde{A}_{2},
\end{equation}
where \(\tilde{A}_{0}\) and \(\tilde{A}_{2}\) are the background-subtracted peak heights of the radial and quadrupole ridges measured from the collapsed profile near the fitted centroids, and selected the \(\Delta\nu_{\mathrm{trial}}\) that maximised \(\mathcal{R}\). The small separation between the radial and quadrupole ridges was required to satisfy
\begin{equation}\label{eq5}
0.07\,\Delta\nu \leq \delta\nu_{02} \leq 0.22\,\Delta\nu,
\end{equation}
consistent with previous studies \citep{Huber2010}. When the quadrupole signature was weak (\(\tilde{A}_{2} < 0.1\,\tilde{A}_{0}\)), we applied a fallback solution using only the radial ridge, which preserves robustness in stars with suppressed quadrupole modes \citep{Stello2016PASA...33...11S}.

\subsection{Precise Parameter Extraction}\label{subsec:final_params}

With the optimal \(\Delta\nu\) determined in Section~\ref{subsec:dnu_estimation}, we refitted the same three-ridge model (\(l=0,1,2\)) to the collapsed \'echelle diagram, now with \(\Delta\nu\) fixed at its optimum. This final fit returns ridge centroids, amplitudes, linewidths, and background, from which we derived the phase shift \(\varepsilon\) and the small frequency separations \(\delta\nu_{02}\) and \(\delta\nu_{01}\) (see panel~(d) of Fig.~\ref{fig:Collapsed}). 

For the radial and quadrupole ridges, we enforced the empirical constraint in Eq.~(\ref{eq5}) to guard against misidentification and overfitting. In some stars, however, the \(l=0\) and \(l=2\) ridges are not cleanly separable in the collapsed \'echelle diagram. If the optimised \(\delta\nu_{02}\) approached either boundary, or if the quadrupole amplitude was negligible (\(A_{l=2} < 0.1\,A_{l=0}\)), we discarded the \(l=2\) ridge and adopted a fallback solution based only on the \(l=0\) ridge. In this case, the radial ridge was still identified robustly, while \(\varepsilon\) remained well defined from its centroid. No analogous fallback treatment was required for the dipole ridge, since the \(l=1\) ridge is generally well separated from the radial ridge.

The dipole-mode small separation was then computed from the fitted \(l=0\) and \(l=1\) centroids as
\begin{equation}
\delta\nu_{01} =
\begin{cases}
\displaystyle \frac{\Delta\nu}{2} - (x_{l=1} - x_{l=0}), & x_{l=1} \geq x_{l=0},\\[4pt]
\displaystyle -\frac{\Delta\nu}{2} - (x_{l=1} - x_{l=0}), & \mathrm{otherwise}.
\end{cases}
\end{equation}
Unlike \(\delta\nu_{02}\), which follows directly from the relative positions of the radial and quadrupole ridges, \(\delta\nu_{01}\) requires this piecewise definition because of the modulo-\(\Delta\nu\) representation of the collapsed diagram. No additional empirical scaling relations (e.g. amplitude ratios or temperature-dependent linewidth trends) were applied at this stage. The final parameters thus follow directly from the constrained fitting and provide an internally consistent solution based on the observed ridge structure in the collapsed diagram.

\section{Sample Selection and Seismic Parameter Extraction}\label{sec:result}

\subsection{Data and Sample Selection}\label{Data}

We used \kepler{} long-cadence photometry covering Quarters Q0-Q17, which provides a nearly continuous 44-month observational baseline. As our sample, we adopted the catalogue of oscillating red giants compiled by \citet{Yu2018}, comprising 16,094 stars identified from the full four-year dataset. We excluded the additional targets reported by \citet{Hon2019MNRAS.485.5616H} that are not in \citet{Yu2018} because, for many of those stars, only \numax{} can be measured robustly from the available light curves. Our analysis and statistics therefore refer exclusively to the \citet{Yu2018} sample.

For all stars in this initial sample, we retrieved the full 4-year set of Pre-search Data Conditioning Simple Aperture Photometry (PDCSAP) light curves \citep{Smith2012PASP..124.1000S,Stumpe2012PASP..124..985S}. Following the methodology outlined in Section~2 of \citet{Sreenivas2024}, we processed the light curves to calculate the PDS. This nuSYD pipeline includes steps for instrumental correction and granulation noise removal, enabling robust extraction of global seismic parameters. As part of this procedure, we also remeasured the frequency of the maximum oscillation power, \numax{}, for each target.

To initialise the large frequency separation, \(\Delta\nu_0\), we used the \Dnu{} values reported by \citet{Yu2018} wherever available (i.e., for all stars in our sample). As a cross-check, we also tested initialisation from the empirical \numax{}–\Dnu{} relation,:
\begin{equation}
\Delta\nu_0 = \alpha \cdot (\nu_{\text{max}})^\beta,
\end{equation}
with \(\alpha = 0.267 \pm 0.002\) and \(\beta = 0.764 \pm 0.002\), as calibrated in \citet{Yu2018} (see their Section~3.6). In practice, the optimisation described in Section~\ref{subsec:dnu_estimation} converged to the same \Dnu{} (within our quoted uncertainties) whether initialised from literature values or from the empirical relation. This robustness reflects the broad ($\pm 8\%$) trial neighbourhood around the initial guess and the fact that the windowing in Section~\ref{Noise Suppression} acts on scales much larger than \Dnu{} and thus does not shift ridge phases.

After combining the available data, we defined a final working set drawn exclusively from the \citet{Yu2018} \kepler{} red-giant sample. Evolutionary states (RGB vs.\ core-helium-burning) were assigned by cross-matching \citet{Vrard2025A&A...697A.165V} to our star list; their catalogue combines six independent seismic diagnostics and reports labels for 18,784 \kepler{} red giants. For stars in our sample without a label in \citet{Vrard2025A&A...697A.165V}, we adopted the classification reported by \citet{Yu2018}; objects lacking a label in both sources were left unclassified.

\begin{figure}
    \centering
    \includegraphics[width=\columnwidth]{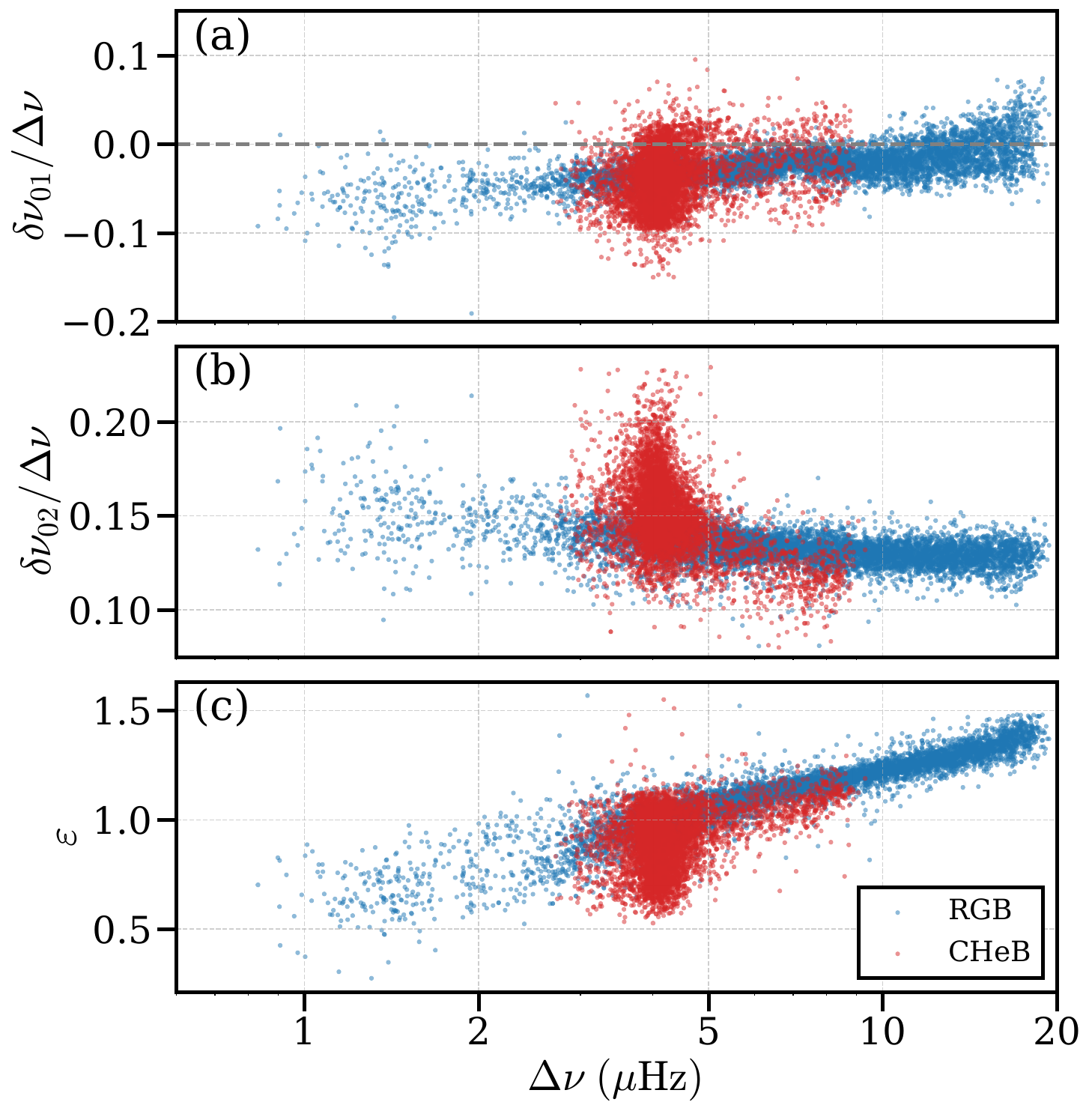}
    \caption{
        Frequency separation ratio diagrams and the $\varepsilon$--$\Delta\nu$ relation.
        (a) $\delta\nu_{01}/\Delta\nu$ versus $\Delta\nu$; (b) $\delta\nu_{02}/\Delta\nu$ versus $\Delta\nu$, and (c) $\varepsilon$ versus $\Delta\nu$.
        Blue and red points represent RGB and CHeB stars, respectively.
    }
    \label{fig:Dnu_three}
\end{figure}

\subsection{Measurements and Overall Statistics}

We determined \(\Delta\nu\), \(\delta\nu_{02}\), \(\delta\nu_{01}\), and \(\varepsilon\) for $16{,}050$, $15{,}279$, $15{,}400$, and $15{,}602$ red giants, respectively. The results are shown in Figure~\ref{fig:Dnu_three} and all measured parameters are listed in Table~\ref{tab:measured_parameters}. The absence of measured values in some stars can be attributed to several factors. First, we did not report a value from when the fit was unstable or the mode identification was uncertain. We flagged a fit as unstable if the optimiser failed to converge or returned a non–positive-definite covariance matrix. We flagged the mode identification as uncertain when the collapsed \'{e}chelle ridge contrast fell below a fixed threshold or when multiple competing maxima of comparable height were present; the formal criteria are given in Section~\ref{subsec:dnu_estimation}. Second, data-related issues, such as poor data quality, insufficient observation durations, or excessive noise, can hinder the detection of the oscillation modes. Finally, astrophysical factors, such as mode suppression, strong magnetic activity/fields, rapid rotation, or mixed-mode crowding, may also play a role, especially in stars that differ significantly from the majority of our sample and might require distinct treatment in terms of mode identification. For example, mode suppression can cause the $l=1$ and $l=2$ modes of oscillation to be absent in some stars \citep{Stello2016Natur.529..364S}. These limitations may lead to the absence of measurements or misidentifications in some stars. 

We estimated uncertainties by generating 200 Monte Carlo realizations per star, perturbing each PDS with a $\chi^2$ distribution with two degrees of freedom, and repeating the full fit. The standard deviation of the resulting parameter samples was adopted as the $1\sigma$ uncertainty \citep{Huber2011}. This PDS-based Monte Carlo procedure has been used extensively in previous asteroseismic analyses of \emph{Kepler} red giants \citep[e.g.][]{Yu2018,Sreenivas2024} and provides a homogeneous set of internal uncertainties across our sample.

In Table~\ref{tab:measured_parameters}, we report the absolute $1\sigma$ uncertainties for the measured quantities $\Delta\nu$, $\varepsilon$, $\delta\nu_{01}$, and $\delta\nu_{02}$, and these same uncertainties are provided in the catalogue so that users can propagate errors directly in the native units of each parameter. We do not provide relative uncertainties of the form $\sigma/\lvert\delta\nu_{0\ell}\rvert$, because $\delta\nu_{01}$ can cross or approach zero, which would make such quantities ill-defined and potentially misleading.

\begin{figure}
    \centering
    \includegraphics[width=\columnwidth]{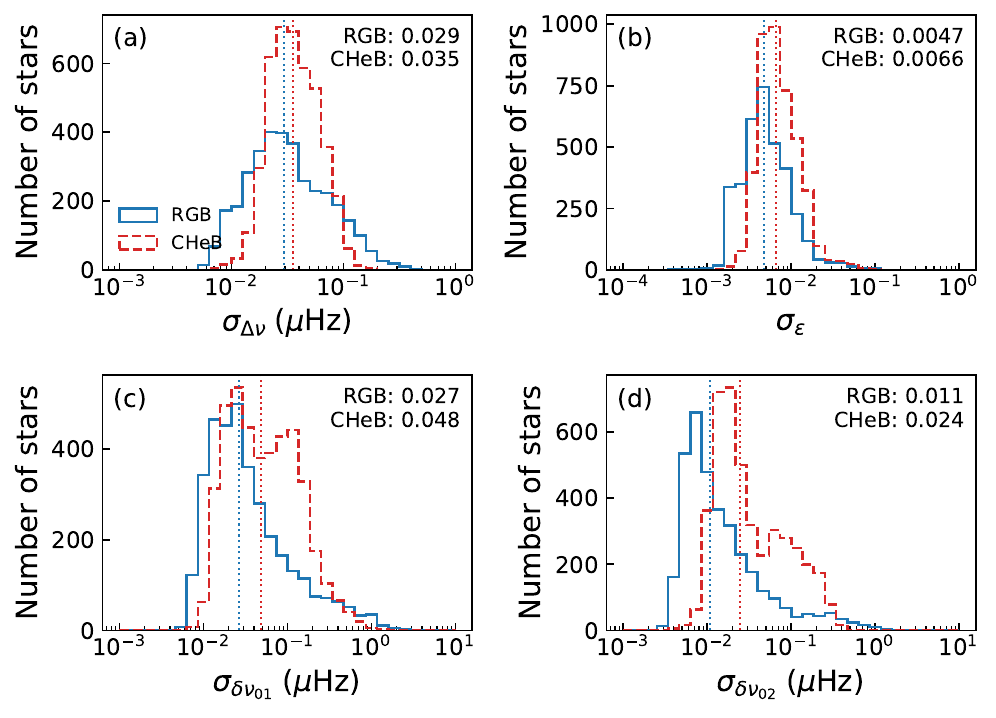}
    \caption{\protect\unboldmath Distributions of the formal $1\sigma$ uncertainties on the seismic parameters. Panels~(a)–(d) show histograms of $\sigma_{\Delta\nu}$, $\sigma_{\varepsilon}$, $\sigma_{\delta\nu_{01}}$, and $\sigma_{\delta\nu_{02}}$ respectively, for RGB (blue) and core-helium-burning (CHeB; red) stars. In each panel, the vertical lines indicate the median uncertainty for each evolutionary phase, and the numerical labels in the upper-right corner list the corresponding median values for the RGB and CHeB samples. The horizontal axis is logarithmic in all panels.}
    \label{fig:uncertainty_hist}
\end{figure}

Figure~\ref{fig:uncertainty_hist} summarises the distributions of these uncertainties for RGB and CHeB stars: the four panels show histograms of $\sigma_{\Delta\nu}$, $\sigma_{\varepsilon}$, $\sigma_{\delta\nu_{01}}$, and $\sigma_{\delta\nu_{02}}$, with RGB and CHeB stars overplotted and median values indicated by vertical lines. The typical (median) uncertainties are ${\sim}\,0.032\,\mu\mathrm{Hz}$ in $\Delta\nu$, ${\sim}\,0.0056$ in $\varepsilon$, ${\sim}\,0.038\,\mu\mathrm{Hz}$ in $\delta\nu_{01}$, and ${\sim}\,0.018\,\mu\mathrm{Hz}$ in $\delta\nu_{02}$.

Finally, we estimated seismic masses for the stars in our sample using the method of \citet{Hon2024ApJ...973..154H}, which is an emulator based on a conditional normalizing flow (CNF). We applied their publicly available code to the global seismic parameters measured in this work, namely \(\Delta\nu\) and \(\nu_{\rm max}\), together with effective temperatures and metallicities where available, to recompute masses for the \emph{Kepler} red giants in our catalogue. These masses, along with the formal uncertainties returned by the Hon et al.\ pipeline, are used in Section~\ref{sec:Mass} to examine mass-dependent trends in the observed seismic diagnostics.

\begin{table}
    \centering
    \caption{Catalogue of measurements for 16,050 red giants. Only the first ten rows are shown here; the complete machine-readable table is available on Zenodo at \href{https://doi.org/10.5281/zenodo.17225960}{10.5281/zenodo.17225960}. Uncertainties in parentheses denote 1$\sigma$.}
    \label{tab:measured_parameters}

    \setlength{\tabcolsep}{3pt}
    \begin{tabular}{@{}lcccc@{}}
        \hline
        KIC & $\Delta\nu$ & $\delta\nu_{01}$ & $\delta\nu_{02}$ & $\varepsilon$ \\
            & ($\mu$Hz)   & ($\mu$Hz)        & ($\mu$Hz)        &               \\
        \hline
        757137 & 3.406(0.019)  & $-0.136$(0.024)  & 0.444(0.015)  & 0.900(0.007) \\
        892010 & 2.451(0.036)  & $-0.113$(0.025)  & 0.383(0.019)  & 0.784(0.013) \\
        892738 & 1.221(0.014)  & $-0.119$(0.010)  & 0.215(0.012)  & 0.958(0.011) \\
        893214 & 4.302(0.010)  & $-0.112$(0.011)  & 0.572(0.007)  & 1.008(0.003) \\
        1026084 & 4.458(0.018)  & $-0.015$(0.013)  & 0.535(0.016)  & 0.950(0.004) \\
        1026180 & 3.967(0.040)  & $-0.198$(0.039)  & 0.629(0.012)  & 0.754(0.008) \\
        1026309 & 1.931(0.044)  & $-0.089$(0.026)  & 0.216(0.013)  & 0.849(0.020) \\
        1026326 & 8.848(0.033)  & $-0.152$(0.073)  & 1.167(0.011)  & 1.172(0.005) \\
        1026452 & 4.023(0.026)  & $-0.037$(0.030)  & 0.668(0.019)  & 0.855(0.006) \\
        1027110 & 1.117(0.013)  & $-0.113$(0.015)  & 0.194(0.016)  & 0.824(0.013) \\
        \hline
    \end{tabular}
\end{table}

\subsubsection{C–D Diagrams and Evolutionary Trends}

Figure~\ref{fig:Dnu_three}(a) shows that \(\delta\nu_{01}\) is negative for most red giants, confirming the findings by \citet{BeddingFirstResultKepler}. We also see a clear trend of the frequency separation ratio \(\delta\nu_{01}/\Delta\nu\) with \(\Delta\nu\): the ratio becomes more negative as \(\Delta\nu\) decreases. This behaviour mirrors, with opposite sign, the trend seen for \(\delta\nu_{02}/\Delta\nu\) and suggests that both small separations respond to similar changes in the internal structure over the same range of \(\Delta\nu\).

Figure~\ref{fig:Dnu_three}(b) shows a modified version of the so-called C--D diagram \citep{Christensen-Dalsgaard1988IAUS..123..295C}, plotting the ratio \(\delta\nu_{02}/\Delta\nu\) versus \(\Delta\nu\). Early \emph{Kepler} ensemble studies presented C--D diagrams for red giants \citep{Huber2010}. Subsequent work showed more clearly that related local seismic diagnostics, including \(\delta\nu_{02}\) and \(\epsilon\), carry information on evolutionary state, and also quantified the mass dependence of \(\delta\nu_{02}\) \citep{Kallinger2012}. C--D diagrams for \emph{Kepler} red giants have also been constructed for cluster stars \citep{Corsaro2012ApJ...757..190C} and for field populations \citep[][their Fig.~A.3]{Miglio2021AA645A85}, where they revealed distinct RGB and RC sequences. The small separation \(\delta\nu_{02}\) measures the frequency difference between two modes with nearly identical eigenfunction shapes in the outer layers of the star, and is sensitive to changes in the slope of the sound speed in the deep interior. For main-sequence stars it tracks the build-up of a helium core and can serve as an age indicator \citep{White2011}, whereas this diagnostic becomes less reliable during the subgiant phase. On the giant branch, the core becomes more centrally concentrated, and theoretical models show that \(\delta\nu_{02}\) respond to the stratification of the hydrogen-burning shell and to the structural differences between RGB and core-helium-burning stars \citep{Montalban2010,Montalban2012ASSP}. As shown in Figure~\ref{fig:Dnu_three}(b), we find that for RGB stars, \(\delta\nu_{02}/\Delta\nu\) varies only weakly with \(\Delta\nu\), with slightly larger values towards lower \(\Delta\nu\), in agreement with previous observational results \citep{BeddingFirstResultKepler,Huber2010}. Meanwhile, CHeB stars display a broader locus in the C--D diagram, a behaviour that we discuss further in Section~\ref{sec:dnu02}.  

We also observe a larger spread in \(\delta\nu_{02}/\Delta\nu\) for stars with \(\Delta\nu < 3 \,\mu\mathrm{Hz}\) than for stars with higher \(\Delta\nu\). This mainly reflects the fact that the plotted quantity is a relative separation: for a given absolute uncertainty in \(\delta\nu_{02}\), the corresponding uncertainty in \(\delta\nu_{02}/\Delta\nu\) becomes larger when \(\Delta\nu\) is smaller.

\subsubsection{\texorpdfstring{\unboldmath The $\varepsilon$--$\Delta\nu$ Relation}{The epsilon--Delta nu Relation}}\label{sec:epsilon}

Figure~\ref{fig:Dnu_three}(c) shows our results for \(\varepsilon\). In contrast to studies based on central estimators, which reported systematic offsets between RGB and CHeB stars \citep{Bedding2011Natur.471..608B,Corsaro2012ApJ...757..190C,Kallinger2012}, our measurements show that the two populations overlap strongly in the \(\Delta\nu\)--\(\varepsilon\) plane, although still with some RGB--CHeB separation. Our best-fitting power-law relation is
\begin{equation}
    \varepsilon = (0.610 \pm 0.002) + (0.625 \pm 0.002) \log_{10}(\Delta\nu /\mu\mathrm{Hz})\,.
\end{equation}
The difference between our results and those of previous studies arises from how both $\Delta\nu$ and the phase offset $\varepsilon$ are defined and measured. In the asymptotic relation for radial modes, $\nu_{n0}\simeq\Delta\nu\,(n+\varepsilon)$, the phase $\varepsilon$ varies slowly with frequency due to curvature from acoustic glitches. Two observational estimators are commonly used: (i) a local, “central'' estimator $\varepsilon_c$ is obtained around $\nu_{\max}$ using the three central radial orders, as in \citet{Kallinger2012} and \citet{Christensen-Dalsgaard2014MNRAS.445.3685C}; (ii) a more global, average estimator $\varepsilon_a$ is the intercept from a least-squares fit of $\nu_{n,0}$ versus $n$ over a broader window, which averages over local modulations \citep[][]{White2011,OngBasu2019ApJ...885...26O}. Because of the way $\varepsilon_c$ is defined, it is particularly sensitive to the curvature and glitch-induced modulation of the radial ridge. As shown by \citet[][their Fig.~9]{Vrard2015A&A...579A..84V}, the radial-mode pattern near $\nu_{\max}$ differs systematically between evolutionary states: CHeB stars often exhibit a characteristic ``C-shape'' in the \'echelle diagram, with $\nu_{\max}$ typically sampling the upper, right-tilted part of the ridge, whereas RGB stars are closer to an ``S-shape''. Using only the three radial orders around $\nu_{\max}$ therefore makes local estimates of $\Delta\nu$ and $\varepsilon$ more sensitive to the radial-ridge morphology in CHeB stars than in RGB stars. This likely contributes to the stronger RGB--CHeB separation reported in $\Delta\nu$--$\varepsilon$ relations based on central estimators. Our analysis, on the other hand, uses five or more radial orders in the collapsed-\'echelle fit, averaging over these glitch-induced perturbations and recovering a more consistent $\varepsilon$--$\Delta\nu$ correlation that is less sensitive to local variations around $\nu_{\max}$.

The presence of acoustic glitches remains important for interpreting the scatter around the mean relation. But, by construction, our method is less sensitive to their local effects and instead emphasises the underlying global trend. A detailed modelling of the glitch signatures themselves is beyond the scope of this paper and could be addressed in future work.

\subsection{Comparison with Existing Catalogues}

\begin{figure}
    \centering
    \includegraphics[width=\columnwidth]{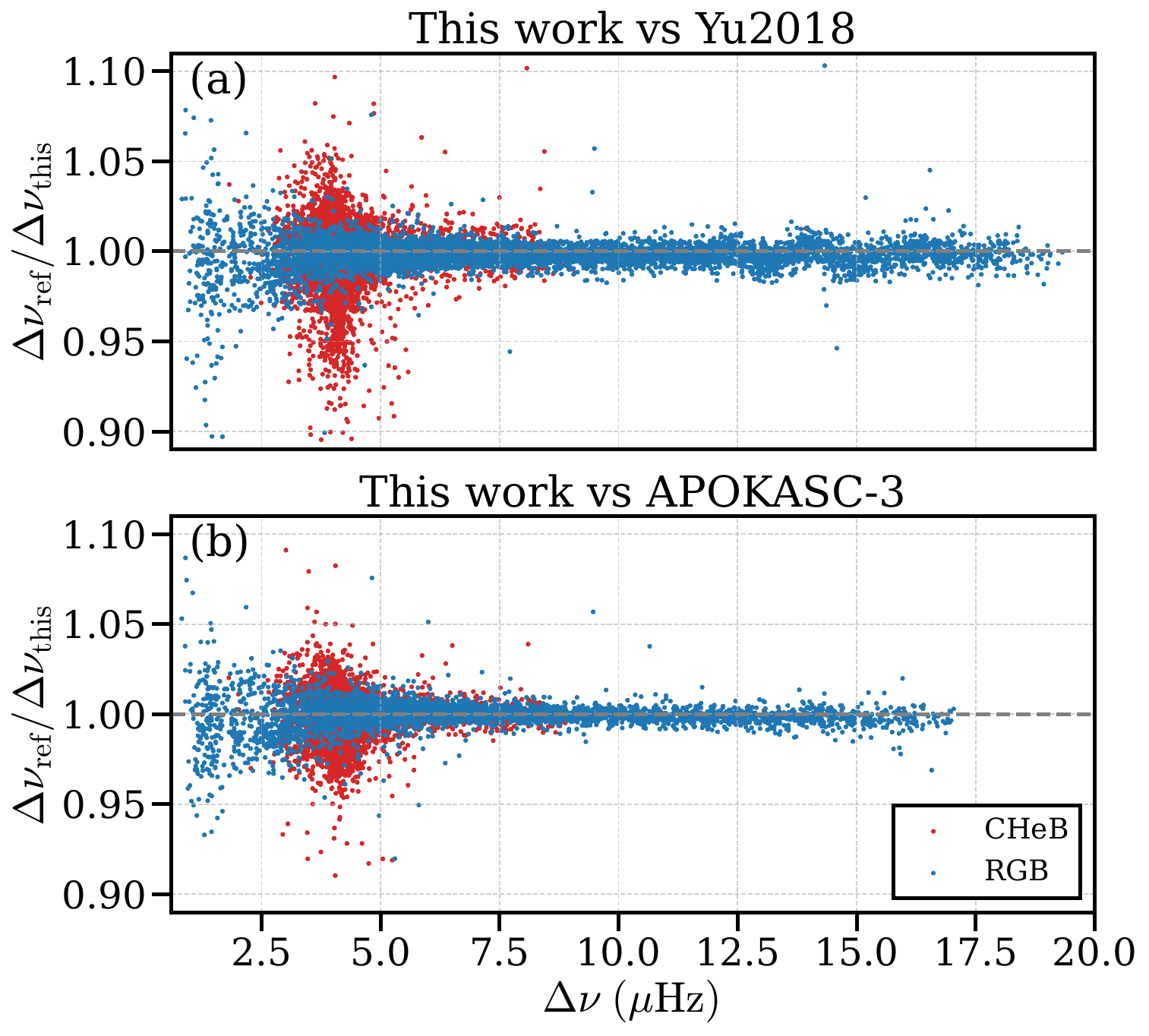}
    \caption{\protect\unboldmath Comparison of \(\Delta\nu\) values derived in this work with those from \citet{Yu2018} and \citet{Pinsonneault2025}. The horizontal axis shows \(\Delta\nu_{\text{this work}}\). (a) The ratio \(\Delta\nu_{\text{Yu18}} / \Delta\nu_{\text{this work}}\), with RGB stars in blue and CHeB stars in red. (b) The ratio \(\Delta\nu_{\text{APOKASC-3}} / \Delta\nu_{\text{this work}}\).}
    \label{fig:Dnu_comparison}
\end{figure}

Figure~\ref{fig:Dnu_comparison} compares the \(\Delta\nu\) values obtained in this work with those from \citet{Yu2018} (upper panel), and with $10{,}079$ stars that overlap with the APOKASC-3 catalogue \citep{Pinsonneault2025} (lower panel), spanning a broad range of evolutionary states. Across the full interval of \(\Delta\nu \in [1,\ 20]\,\mu\mathrm{Hz}\), our measurements show excellent consistency with APOKASC-3, which was not used as input to our pipeline and therefore provides an external benchmark for our \(\Delta\nu\) measurements. However, compared to the SYD pipeline results from \citet{Yu2018}, we observe an oscillatory structure in the residuals for stars with \(\Delta\nu > 11\,\mu\mathrm{Hz}\). This pattern reflects a subtle systematic error in SYD measurements. Since the Yu et al.\ (2018) values are used in our analysis as initial guesses and to define the search ranges for \(\Delta\nu\), this comparison should be regarded as an internal consistency check rather than a fully independent validation. Despite this, over \(99\%\) of the RGB sample lies within a \(1\%\) deviation, indicating strong overall agreement. 

For core helium-burning (CHeB) stars (red points in Figure~\ref{fig:Dnu_comparison}), a small systematic offset is evident relative to both APOKASC-3 and \citet{Yu2018}. This discrepancy primarily arises from oscillation glitches that are characteristic of the HeB phase \citep{Vrard2015A&A...579A..84V}. Our clipping approach (Section~\ref{step:prep_spectrum}), which attenuates anomalously strong peaks, tends to yield slightly lower \(\Delta\nu\) values and correspondingly higher \(\varepsilon\) estimates. A discussion of how glitch structures and peak clipping affect the inference of \(\Delta\nu\) and \(\varepsilon\) is presented in Section~\ref{sec:epsilon}.

As an external check, we cross-matched our catalogue with the APOKASC peak-bagging released by \citet{Kallinger2019arXiv190609428K}, and compared the quadrupole small separation on a star-by-star basis. We found no evidence for a systematic offset in \(\delta\nu_{02}\) between the two catalogues; the relative differences are small and consistent with our uncertainties. This agreement supports the validity of the collapsed-\'{e}chelle measurements adopted here. A direct comparison of \(\varepsilon\) is not presented because the Kallinger release provides central three-order quantities tied to local definitions, whereas our work reports a global (asymptotic) estimator; mixing these two conventions would not constitute a like-for-like test (see discussion in Section~\ref{sec:epsilon}).

\section{Comparison with models}\label{sec:Mass}

\subsection{Stellar Models}

To compare our results with theoretical predictions, we computed a set of stellar evolutionary models using MESA \citep[r24.03.1][]{mesa2011,mesa2013,mesa2015,mesa2018,mesa2019,mesa2023} and GYRE \citep[v7.1][]{gyre2013} implemented within \texttt{run\_star\_extras} \citep[GYRE on-the-fly;][]{Bellinger2022, Joyce++2024}. Evolutionary tracks were calculated for stellar masses ranging from 0.6 to 3.6\,M$_\odot$ in steps of 0.2\,M$_\odot$, and for initial metallicities from $-1.2$ to 0.4\,dex in steps of 0.4\,dex. The initial helium abundance varied with metal abundance linearly according to the relation $Y_{\rm init} = 0.249 + 1.5Z_{\rm init}$ \citep{Planck2016,Choi2016}. The solar abundance scale follows \citet{agss09}, with $Z_{\odot} = 0.0134$ and $X_{\odot} = 0.7381$. The mixing length parameter was fixed at 2.2. The convective core overshoot was modelled as a function of stellar mass, using the fitting relation from \citet{Claret2019}. Other mixing processes, including convective shell overshoot, semiconvection, and thermohaline mixing, were set according to the MIST isochrone settings \citep{Choi2016}. The full MESA working directory, including inlist files and other custom settings used to generate these models, is publicly available on Zenodo at \href{https://doi.org/10.5281/zenodo.17226468}{10.5281/zenodo.17226468}.

For each model, we computed oscillation frequencies for modes with spherical degrees $0 \leq l \leq 2$ in the frequency range around \numax{}. For non-radial modes, we included solutions from both the full set of oscillation equations, which yield mixed modes \citep{gyre2013}, and the reduced set in the $\omega^2 \gg N^2$ limit ($\omega$ is the angular frequency and $N$ is the buoyancy frequency). The latter allows computation of $\pi$ modes, which are pure p modes in the absence of g-mode coupling \citep{Ong2020}.

To place the models on the same footing as the observations, each MESA+GYRE snapshot was converted into a synthetic PDS centred on \numax. We placed a Lorentzian at every $l=0,1,2$ mode frequency, modulated by a Gaussian envelope
$\propto \exp[-(\nu-\numax)^2/(2\sigma^2)]$ with $\sigma = 0.66\,(\numax/\mu\mathrm{Hz})^{0.888}/2.355$ \citep{Mosser2012A&A...537A..30M}, and adopted fixed relative mode visibilities of $A_{l=0}\!:\!A_{l=1}\!:\!A_{l=2}=1:0.8:0.6$. Line widths were prescribed as simple functions of $\Delta\nu$ and evolutionary state (RGB vs.\ CHeB). The PDS was lightly smoothed and collapsed into an échelle spectrum using exactly the same settings as for the observations.

\begin{figure*}
    \centering
    \includegraphics[width=0.9\textwidth]{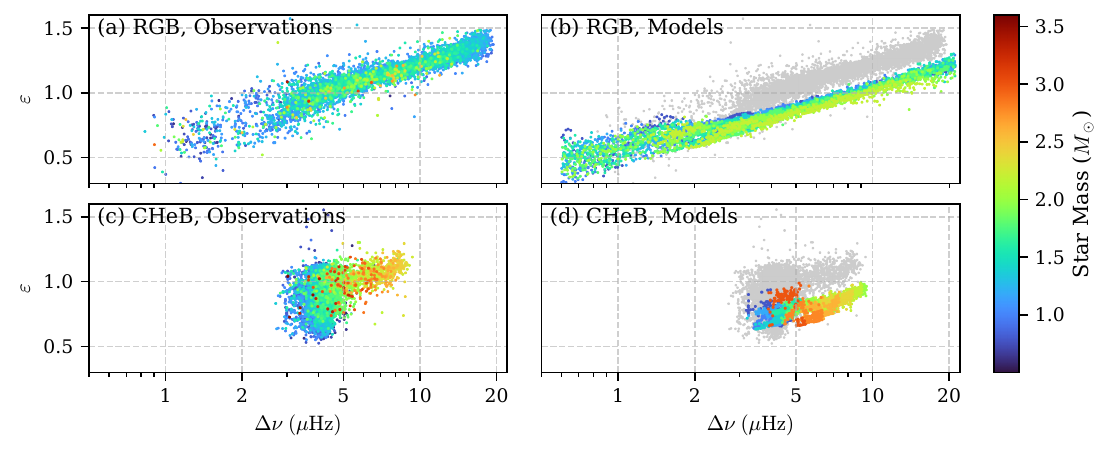}
    \caption{\protect\unboldmath Comparison between $\varepsilon$ and $\Delta\nu$ for observations and models, colour-coded by stellar mass.
    Panels (a) and (b) show RGB stars, while panels (c) and (d) show CHeB stars.
    In each row, the left panel displays the observed values and the right panel shows the corresponding $\pi$-mode model predictions; in the right-hand panels, grey points in the background reproduce the observations for reference.}
    \label{fig:model_eps}
\end{figure*}

\begin{figure*}
    \centering
    \includegraphics[width=0.9\textwidth]{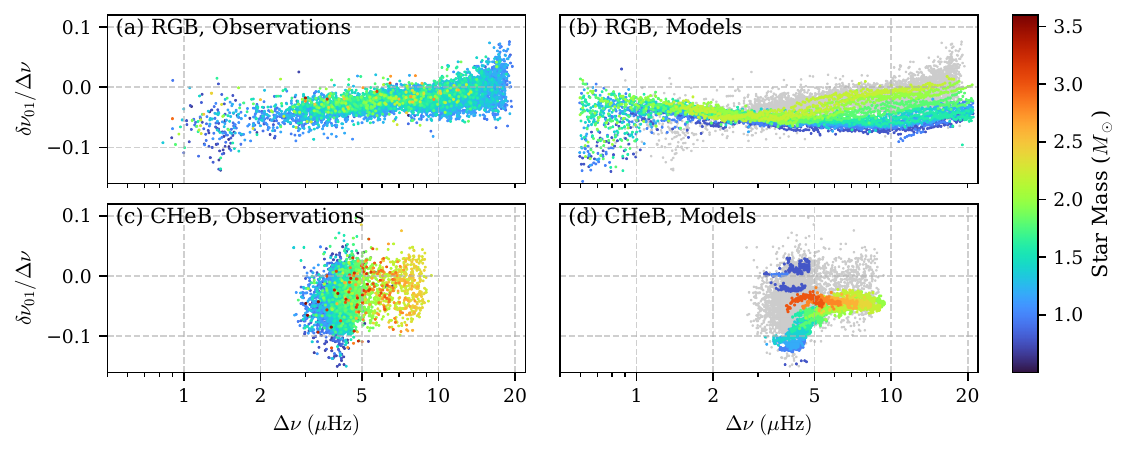}
    \caption{Same as Figure~\ref{fig:model_eps}, but for \protect\unboldmath $\delta\nu_{01}$.}
    \label{fig:model_dnu01}
\end{figure*}

\begin{figure*}
    \centering
    \includegraphics[width=0.9\textwidth]{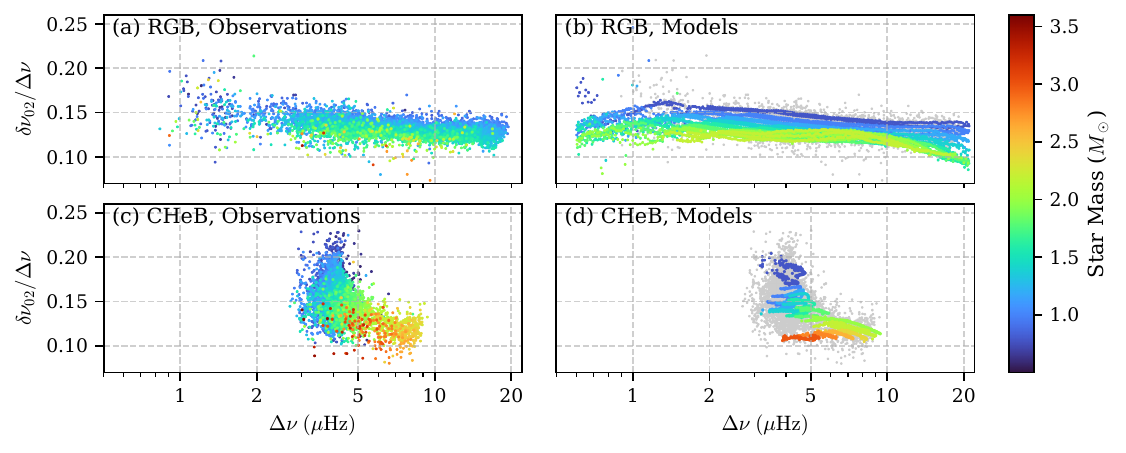}
    \caption{Same as Figure~\ref{fig:model_eps}, but for \protect\unboldmath $\delta\nu_{02}$.}
    \label{fig:model_dnu02}
\end{figure*}

We measured $\Delta\nu$, $\varepsilon$, $\delta\nu_{01}$, and $\delta\nu_{02}$ from each synthetic PDS using the same pipeline as for the observations. For direct comparison we also report the dimensionless ratios $\delta\nu_{01}/\Delta\nu$ and $\delta\nu_{02}/\Delta\nu$. Note that there are not the same as the \(r_{01}\) and \(r_{02}\) ratios used by \cite{Roxburgh2003A&A...411..215R} and others, which are based on order-by-order ratios of modes, whereas \(\delta\nu_{01}\), \(\delta\nu_{02}\) and \(\Delta\nu\) are averaged over all orders. We summarise the comparison between observations and models in three figures (Figures~\ref{fig:model_eps}--\ref{fig:model_dnu02}), which show each seismic parameter as a function of $\Delta\nu$ for both RGB and CHeB stars. The model points shown were measured on the synthetic PDS using the procedure above, ensuring that $\Delta\nu$, $\delta\nu_{01}$, $\delta\nu_{02}$, and $\varepsilon$ are defined identically for models and observations.

\subsection{\texorpdfstring{\unboldmath $\varepsilon$ \& $\delta\nu_{01}$: Surface Corrections}
                        {epsilon and delta nu 01: Surface Corrections}}

Figure~\ref{fig:model_eps} compares the observed \( \varepsilon \) values with those predicted by stellar models. A systematic deviation is evident across both RGB and CHeB stars, indicating that current models do not fully capture the physics in the outer stellar layers. The linear relationship between \( \varepsilon \) and \( \Delta\nu \) (Equation~7) shows an offset between models and observations, which is consistent with the findings of \citet{White2011}. This behaviour is naturally interpreted as the signature of near-surface modelling uncertainties, including turbulent convection, atmospheric boundary conditions, and non-adiabatic effects. The similarity of the offset between RGB and CHeB stars suggests that, at least to first order, a single prescription for the near-surface contribution might be applicable across both evolutionary phases. This result can be combined with the findings of \citet{Li2023MNRAS.523..916L}, who derived a surface-correction formula for RGB stars based on surface gravity, effective temperature, and metallicity. Our data suggest that their prescription is qualitatively compatible with the systematic offsets in \(\varepsilon\) observed in both RGB and CHeB stars, hinting at the possibility of a unified surface-correction framework for evolved stars. A full recalibration of such a framework, however, is beyond the scope of this paper; see also \citet{Schimak2026MNRAS.546ag151S} for a recent discussion of surface corrections and phase-shift offsets in red-clump modelling.

Figure~\ref{fig:model_dnu01} shows clear differences between the observed and modelled \(\delta\nu_{01}\) values, especially for RGB stars, whereas the measured \(\delta\nu_{01}\) for CHeB stars shows a wide spread. The latter is likely because their high coupling strength makes $l=1$ mixed modes difficult to identify \citep{Dhanpal2023ApJ...958...63D}, adding measurement uncertainty to \( \delta\nu_{01} \). For RGB stars, the systematic offset indicates that the same set of models that qualitatively reproduces the \(\varepsilon\)--\(\Delta\nu\) relation still does not capture all of the physics that shapes the $l=1$ p-mode pattern.

The presence of systematic differences in both \eps{} and \dnuol{} therefore highlights the sensitivity of these diagnostics to the outer-envelope and near-surface structure in evolved stars. However, several physical ingredients can, in principle, contribute to the RGB discrepancy in \dnuol{}, including the treatment of near-surface convection, the atmospheric boundary, the stratification of the outer envelope, and the coupling between p and g modes. In the present work we do not explore different surface-correction prescriptions or variations in the interior physics, and so we cannot uniquely attribute the RGB offset in \dnuol{} to any single physical ingredient (e.g. the surface term) alone.

Future modelling efforts that explicitly vary the surface term and the outer-envelope structure --- for example by applying generalised surface corrections to models of evolved stars \citep{Ball2014A&A...568A.123B,Ball2017A&A...600A.128B} --- will be required to test whether a single prescription can simultaneously reproduce both \eps{} and \dnuol{} for RGB and CHeB stars. In particular, \citet{Ball2018MNRAS.478.4697B} have already investigated surface terms for \emph{Kepler} RGB stars using such formulations, providing a template for future applications to larger evolved-star samples. Our results emphasise that small differences involving $l=1$ modes provide a stringent observational constraint on these developments, but a detailed calibration of mode-dependent or mixed-mode-specific surface corrections lies beyond the scope of this catalogue paper.

\subsection[delta nu 02: A Strong Constraint for Both RGB and CHeB]
{\texorpdfstring{$\delta\nu_{02}$: A Strong Constraint for Both RGB and CHeB}
                {delta nu 02: A Strong Constraint for Both RGB and CHeB}}
\label{sec:dnu02}

For RGB stars, Figure~\ref{fig:model_dnu02}(a) confirms that \( \delta\nu_{02}\) is proportional to \( \Delta\nu\), but with a clear mass dependence. This aligns with previous studies \citep{Huber2010,Handberg2017}, which established that \( \delta \nu_{02} \) remains nearly constant as a fraction of \( \Delta \nu \) during the RGB stage. The observed data match theoretical predictions from the MESA and GYRE models, confirming that \( \delta \nu_{02} \) can constrain stellar masses. However, its diagnostic power for RGB stars is limited, as it primarily reflects global seismic properties rather than detailed internal structure. This constancy arises because the radiative core of an RGB star evolves minimally in size, making \(\delta\nu_{02}\) a stable but less informative parameter compared to \(\Delta\nu\) or \numax.

In contrast, \( \delta \nu_{02}\) in CHeB stars exhibits a more complex behaviour (Figure~\ref{fig:model_dnu02}(c)). A clear separation emerges, splitting stars into two distinct sequences: red clump (RC) and secondary clump (SC). This separation reflects differences in core helium ignition processes, which are mass-dependent. Low-mass RC stars undergo a helium flash at the RGB tip, creating a dense, stratified core. Higher-mass SC stars ignite helium under non-degenerate conditions, resulting in a smoother core structure (see review by \citealt{Girardi2016ARA&A..54...95G}).

The ratio \( \delta\nu_{02}/\Delta\nu\) serves as a diagnostic for these structural differences. The mass- and evolutionary-state dependence of $\delta\nu_{02}/\Delta\nu$ was anticipated by stellar-model calculations \citep[][see their Fig.~4]{Montalban2010ApJL721L182}.
An extended discussion of the underlying physics and its diagnostic use in red giants is provided by \citet{Montalban2012ASSP}. Recent TESS ensemble analysis also reports a pronounced increase of $\delta\nu_{02}/\Delta\nu$ toward lower $\Delta\nu$, and a weaker correlation but stronger mass dependence at $\Delta\nu\gtrsim15.6\,\mu\mathrm{Hz}$ \citep{Zhou2025ApJS27937}. For RC stars, the ratio spans a relatively wide range from approximately 0.12 to 0.23 in Figure~\ref{fig:model_dnu02}(c), exhibiting a clear inverse correlation with stellar mass: more massive red clump stars tend to have lower values of \(\delta\nu_{02}/\Delta\nu\). This trend suggests that even within the red clump, structural variations determined by mass—such as differences in core mass and envelope stratification—significantly impact the small separation.

\begin{figure*}
\centering
\includegraphics[width=0.9\textwidth]{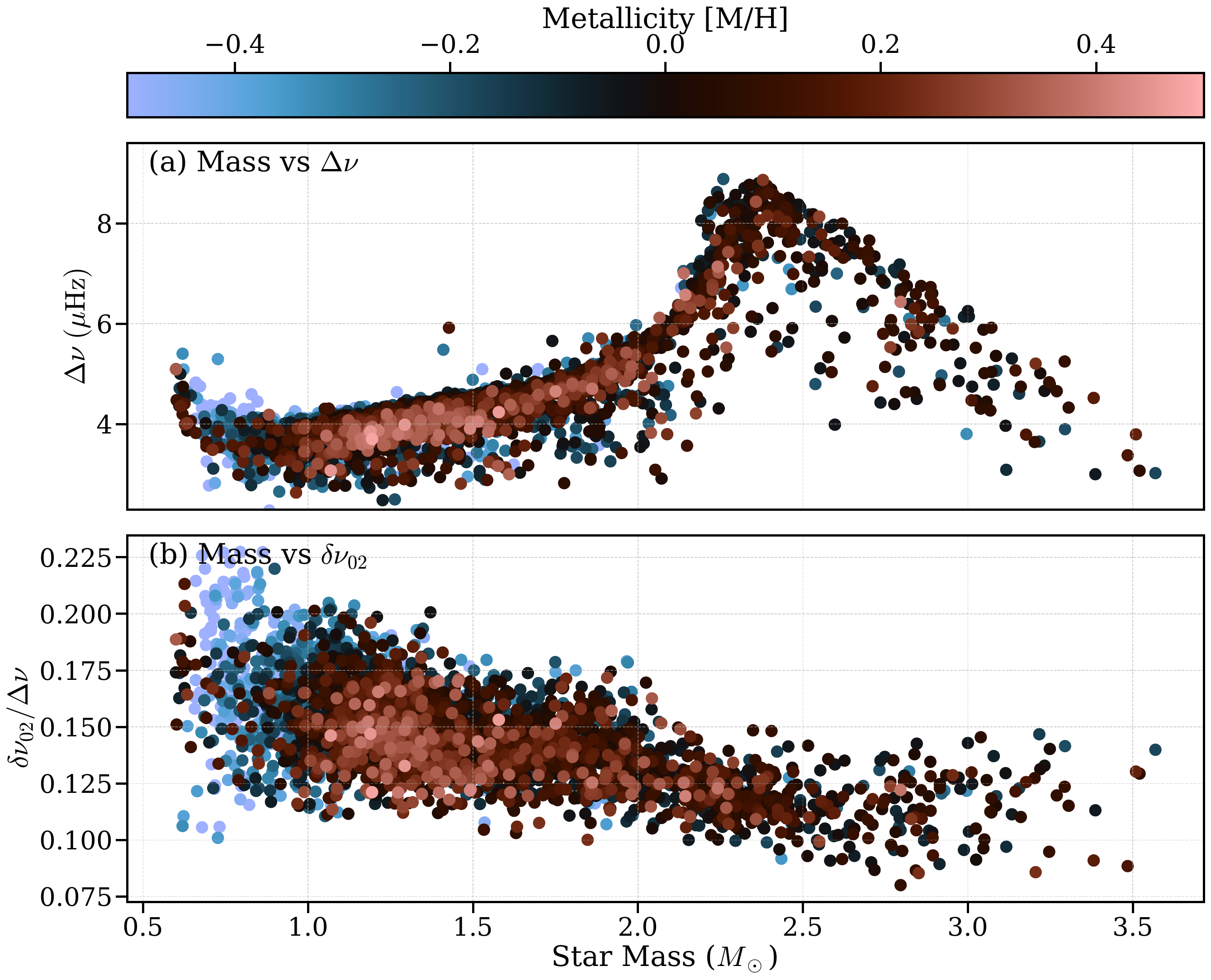}
\caption{\protect\unboldmath Core–He–burning stars with APOKASC-3 metallicities: (a) mass versus \Dnu; (b) mass versus \(\delta\nu_{02}/\Delta\nu\). Points are coloured by \([\mathrm{M/H}]\).}
\label{fig:mass_meta}
\end{figure*}

In contrast, SC stars display a narrow distribution around \( \delta\nu_{02}/\Delta\nu \sim 0.10\), with little apparent mass dependence. This mass-independent clustering is consistently reproduced by the theoretical models (Figure~\ref{fig:model_dnu02}(d)), where both observations and models show little spread. The weak sensitivity of \(\delta\nu_{02}/\Delta\nu\) to mass in this regime is consistent with the more gradual evolutionary transitions these stars experience: without a helium flash, the build-up of the core–envelope density contrast is smoother, so that \(l=2\) modes remain predominantly p-dominated. Consequently, \(\delta\nu_{02}/\Delta\nu\) exhibits a tighter clustering near \(\sim0.10\) \citep[e.g.][]{Girardi2016ARA&A..54...95G,Hekker2017A&ARv..25....1H}.

Taken together, these patterns show that \(\delta\nu_{02}\) is a powerful seismic diagnostic of the structural differences between RC and SC stars, with the two populations occupying distinct, though partially overlapping, sequences in the \(\Delta\nu-\delta\nu_{02}/\Delta\nu\) plane. The sensitivity of \(\delta\nu_{02}\) to mass in RC stars further informs calibrations of core mixing processes and convective boundary treatments \citep{Ong2025ApJ...980..199O}. For SC stars, the uniformity of \(\delta\nu_{02}/\Delta\nu\) suggests that helium-burning efficiency dominates over structural variations. Together, these results highlight \(\delta\nu_{02}\)
 as a critical probe of stellar evolution, complementing \Dnu{} and \numax{} in constraining internal structure.

\subsection{Metallicity and the Helium-Flash Transition in Core He-Burning Stars}\label{subsec:meta_heflash}

To test whether composition modulates the structure of CHeB stars, we examined the relations between stellar mass and the seismic parameters \Dnu{} and \dnuoz{} for stars classified as CHeB. These relations are shown in Figure~\ref{fig:mass_meta}, colour-coded by spectroscopic metallicity \([\mathrm{M/H}]\) from APOKASC-3 \citep[][]{Pinsonneault2025}. Not all stars in our catalogue have spectroscopic metallicities; the CHeB subsample with available \([\mathrm{M/H}]\) contains 5063 stars.

Figure~\ref{fig:mass_meta} reveals three qualitative features. First, in the mass-\Dnu{} plane there is a sharp peak in the range 2--2.3\,$M_\odot$, which corresponds to the transition between RC and SC \citep[e.g.][]{Girardi1999MNRAS.308..818G,Girardi2000A&AS..141..371G,Bressan2012MNRAS.427..127B,Constantino2015MNRAS.452..123C}. Second, the mass--\dnuoz{} relation shows a similar transition, broadly mirroring the trend in mass--\Dnu{}, as in Figure~\ref{fig:model_dnu02}(c). Third, at fixed mass both panels display a colour gradient, suggesting that \([\mathrm{M/H}]\) influences either the location or the sharpness of the helium flash limit peak. This is qualitatively expected: in stellar models, the helium-flash threshold mass ($M_{\mathrm{HeF}}$) depends on both composition and the treatment of convective-boundary mixing, because opacity and core growth regulate whether helium ignites under partial degeneracy.

These patterns are empirical indications rather than a calibrated relation. A quantitative test of [M/H] vs. $M_{\mathrm{HeF}}$ sensitivity would require a larger CHeB sample with spectroscopic \([\mathrm{M/H}]\) and uniform mass determinations, together with a grid that varies $Z$ and envelope overshoot in a controlled way \citep[e.g.][]{Bossini2015MNRAS.453.2290B,Chaplin2013ARA&A..51..353C}. Future work could extend this analysis with TESS targets to increase the number of CHeB stars with \([\mathrm{M/H}]\), enabling a direct measurement of how metallicity shapes the observed mass–\Dnu{} and mass–\dnuoz{} sequences and the inferred helium–flash limit.

\section{Conclusions}\label{sec:Conclusions}

We have presented a homogeneous analysis of $\sim$16,000 red giants observed by \kepler{} using all available long-cadence data sets. Our work provides a comprehensive catalogue of global seismic parameters and investigates their evolutionary dependencies. The main results are summarized as follows:
\begin{enumerate}

\item We provide a homogeneous catalogue of $\Delta\nu$, $\delta\nu_{01}$, $\delta\nu_{02}$, and $\varepsilon$ for $16{,}050$ red giants, representing the largest sample to date for these parameters. The typical (median) uncertainties are $0.3\%$ in $\Delta\nu$, $0.0008$ in $\delta\nu_{01}/\Delta\nu$, $0.0010$ in $\delta\nu_{02}/\Delta\nu$, and $0.3\%$ in $\varepsilon$.

\item The ratio $\delta\nu_{02}/\Delta\nu$ is nearly constant for RGB stars, consistent with theoretical models. In contrast, CHeB stars exhibit a clear separation into RC and SC sequences. This distinction reflects differences in core helium ignition processes, with $\delta\nu_{02}/\Delta\nu$ serving as a seismic diagnostic for CHeB sub-populations.

\item We derived a unified power-law relation $\varepsilon = 0.610 + 0.625\log_{10}(\Delta\nu)$ for all stars. Systematic offsets between observations and models indicate the effects of a poorly modelled surface (known as the "surface effect"), which are consistent across both the RGB and CHeB phases. This suggests a single surface correction prescription may apply to both types of stars.

\item Observed $\delta\nu_{01}$ values are systematically offset from models, with negative values for most red giants. The persistence of offsets in models after uniform surface corrections implies mode-dependent surface effects, suggesting separate corrections for $l=1$ modes.

\item For RC stars, $\delta\nu_{02}/\Delta\nu$ correlates inversely with mass, providing a constraint on core structure and mixing processes. SC stars show uniform $\delta\nu_{02}/\Delta\nu$ values, reflecting smoother core-envelope density contrasts. This highlights $\delta\nu_{02}$ as a critical probe of stellar evolution, complementing $\Delta\nu$ and $\nu_{\text{max}}$.
\end{enumerate}

These results underscore the value of small frequency separations and phase shifts in constraining stellar interiors. The $\delta\nu_{02}$ parameter, in particular, emerges as a powerful tool for distinguishing evolutionary states and probing core helium ignition processes. These measurements will allow others to leverage these parameters to refine stellar models, particularly in the treatment of surface-term corrections and convective-boundary mixing.

\section*{Acknowledgements}

We acknowledge support from the Australian Research Council through Laureate Fellowship FL220100117. 
DH acknowledges support from the Alfred P. Sloan Foundation, the National Aeronautics and Space Administration (80NSSC19K0597 and 80NSSC21K0652) and the Australian Research Council (FT200100871).

%%%%%%%%%%%%%%%%%%%%%%%%%%%%%%%%%%%%%%%%%%%%%%%%%%
\section*{Data availability}

The full catalogue underlying this article, containing all measured values and 1$\sigma$ uncertainties for $\Delta\nu$, $\delta\nu_{01}$, $\delta\nu_{02}$, and $\varepsilon$, is available on Zenodo at \href{https://doi.org/10.5281/zenodo.17225960}{10.5281/zenodo.17225960}. The MESA models used in this work are also available as a Parquet file on Zenodo at \href{https://doi.org/10.5281/zenodo.17226468}{10.5281/zenodo.17226468}. The catalogue will also be made available through CDS/VizieR. The \kepler\ light curves analysed in this work are publicly available from the MAST Portal at \url{https://mast.stsci.edu/portal/Mashup/Clients/Mast/Portal.html}.



\bibliographystyle{mnras}
\bibliography{reference} % if your bibtex file is called example.bib

@ARTICLE{Joyce++2024,
       author = {{Joyce}, Meridith and {Moln{\'a}r}, L{\'a}szl{\'o} and {Cinquegrana}, Giulia and {Karakas}, Amanda and {Tayar}, Jamie and {Tarczay-Neh{\'e}z}, D{\'o}ra},
        title = "{Stellar Evolution in Real Time. II. R Hydrae and an Open-Source Grid of >3000 Seismic TP-AGB Models Computed with MESA}",
      journal = {\apj},
         year = 2024,
        month = aug,
       volume = {971},
       number = {2},
          eid = {186},
        pages = {186},
          doi = {10.3847/1538-4357/ad534a},
archivePrefix = {arXiv},
       eprint = {2401.16142},
 primaryClass = {astro-ph.SR},
       adsurl = {https://ui.adsabs.harvard.edu/abs/2024ApJ...971..186J}
}

@ARTICLE{Bellinger2022,
       author = {{Bellinger}, Earl Patrick and {Christensen-Dalsgaard}, J{\o}rgen},
        title = "{Towards solar measurements of nuclear reaction rates}",
      journal = {\mnras},
         year = 2022,
        month = dec,
       volume = {517},
       number = {4},
        pages = {5281-5288},
          doi = {10.1093/mnras/stac1845},
archivePrefix = {arXiv},
       eprint = {2206.13570},
 primaryClass = {astro-ph.SR},
       adsurl = {https://ui.adsabs.harvard.edu/abs/2022MNRAS.517.5281B}
}

@ARTICLE{Planck2016,
       author = {{Planck Collaboration} and {Ade}, P.~A.~R. and {Aghanim}, N. and {Arnaud}, M. and {Ashdown}, M. and {Aumont}, J. and {Baccigalupi}, C. and {Banday}, A.~J. and {Barreiro}, R.~B. and {Bartlett}, J.~G. and et al.},
        title = "{Planck 2015 results. XIII. Cosmological parameters}",
      journal = {\aap},
         year = 2016,
        month = sep,
       volume = {594},
          eid = {A13},
        pages = {A13},
          doi = {10.1051/0004-6361/201525830},
archivePrefix = {arXiv},
       eprint = {1502.01589},
 primaryClass = {astro-ph.CO},
       adsurl = {https://ui.adsabs.harvard.edu/abs/2016A&A...594A..13P}
}

@ARTICLE{Ong2020,
       author = {{Ong}, J.~M. Joel and {Basu}, Sarbani},
        title = "{Semianalytic Expressions for the Isolation and Coupling of Mixed Modes}",
      journal = {\apj},
         year = 2020,
        month = aug,
       volume = {898},
       number = {2},
          eid = {127},
        pages = {127},
          doi = {10.3847/1538-4357/ab9ffb},
archivePrefix = {arXiv},
       eprint = {2006.13313},
 primaryClass = {astro-ph.SR},
       adsurl = {https://ui.adsabs.harvard.edu/abs/2020ApJ...898..127O}
}

@ARTICLE{Choi2016,
       author = {{Choi}, Jieun and {Dotter}, Aaron and {Conroy}, Charlie and {Cantiello}, Matteo and {Paxton}, Bill and {Johnson}, Benjamin D.},
        title = "{Mesa Isochrones and Stellar Tracks (MIST). I. Solar-scaled Models}",
      journal = {\apj},
         year = 2016,
        month = jun,
       volume = {823},
       number = {2},
          eid = {102},
        pages = {102},
          doi = {10.3847/0004-637X/823/2/102},
archivePrefix = {arXiv},
       eprint = {1604.08592},
 primaryClass = {astro-ph.SR},
       adsurl = {https://ui.adsabs.harvard.edu/abs/2016ApJ...823..102C}
}

@ARTICLE{Claret2019,
       author = {{Claret}, Antonio and {Torres}, Guillermo},
        title = "{The Dependence of Convective Core Overshooting on Stellar Mass: Reality Check and Additional Evidence}",
      journal = {\apj},
         year = 2019,
        month = may,
       volume = {876},
       number = {2},
          eid = {134},
        pages = {134},
          doi = {10.3847/1538-4357/ab1589},
archivePrefix = {arXiv},
       eprint = {1904.02714},
 primaryClass = {astro-ph.SR},
       adsurl = {https://ui.adsabs.harvard.edu/abs/2019ApJ...876..134C}
}

@ARTICLE{agss09,
       author = {{Asplund}, Martin and {Grevesse}, Nicolas and {Sauval}, A. Jacques and {Scott}, Pat},
        title = "{The Chemical Composition of the Sun}",
      journal = {\araa},
         year = 2009,
        month = sep,
       volume = {47},
       number = {1},
        pages = {481-522},
          doi = {10.1146/annurev.astro.46.060407.145222},
archivePrefix = {arXiv},
       eprint = {0909.0948},
 primaryClass = {astro-ph.SR},
       adsurl = {https://ui.adsabs.harvard.edu/abs/2009ARA&A..47..481A}
}

@ARTICLE{mesa2023,
       author = {{Jermyn}, Adam S. and {Bauer}, Evan B. and {Schwab}, Josiah and {Farmer}, R. and {Ball}, Warrick H. and {Bellinger}, Earl P. and {Dotter}, Aaron and {Joyce}, Meridith and {Marchant}, Pablo and {Mombarg}, Joey S.~G. and {Wolf}, William M. and {Sunny Wong}, Tin Long and {Cinquegrana}, Giulia C. and {Farrell}, Eoin and {Smolec}, R. and {Thoul}, Anne and {Cantiello}, Matteo and {Herwig}, Falk and {Toloza}, Odette and {Bildsten}, Lars and {Townsend}, Richard H.~D. and {Timmes}, F.~X.},
        title = "{Modules for Experiments in Stellar Astrophysics (MESA): Time-dependent Convection, Energy Conservation, Automatic Differentiation, and Infrastructure}",
      journal = {\apjs},
         year = 2023,
        month = mar,
       volume = {265},
       number = {1},
          eid = {15},
        pages = {15},
          doi = {10.3847/1538-4365/acae8d},
archivePrefix = {arXiv},
       eprint = {2208.03651},
 primaryClass = {astro-ph.SR},
       adsurl = {https://ui.adsabs.harvard.edu/abs/2023ApJS..265...15J}
}

@ARTICLE{mesa2019,
       author = {{Paxton}, Bill and {Smolec}, R. and {Schwab}, Josiah and {Gautschy}, A. and {Bildsten}, Lars and {Cantiello}, Matteo and {Dotter}, Aaron and {Farmer}, R. and {Goldberg}, Jared A. and {Jermyn}, Adam S. and {Kanbur}, S.~M. and {Marchant}, Pablo and {Thoul}, Anne and {Townsend}, Richard H.~D. and {Wolf}, William M. and {Zhang}, Michael and {Timmes}, F.~X.},
        title = "{Modules for Experiments in Stellar Astrophysics (MESA): Pulsating Variable Stars, Rotation, Convective Boundaries, and Energy Conservation}",
      journal = {\apjs},
         year = 2019,
        month = jul,
       volume = {243},
       number = {1},
          eid = {10},
        pages = {10},
          doi = {10.3847/1538-4365/ab2241},
archivePrefix = {arXiv},
       eprint = {1903.01426},
 primaryClass = {astro-ph.SR},
       adsurl = {https://ui.adsabs.harvard.edu/abs/2019ApJS..243...10P}
}

@ARTICLE{mesa2018,
       author = {{Paxton}, Bill and {Schwab}, Josiah and {Bauer}, Evan B. and {Bildsten}, Lars and {Blinnikov}, Sergei and {Duffell}, Paul and {Farmer}, R. and {Goldberg}, Jared A. and {Marchant}, Pablo and {Sorokina}, Elena and {Thoul}, Anne and {Townsend}, Richard H.~D. and {Timmes}, F.~X.},
        title = "{Modules for Experiments in Stellar Astrophysics (MESA): Convective Boundaries, Element Diffusion, and Massive Star Explosions}",
      journal = {\apjs},
         year = 2018,
        month = feb,
       volume = {234},
       number = {2},
          eid = {34},
        pages = {34},
          doi = {10.3847/1538-4365/aaa5a8},
archivePrefix = {arXiv},
       eprint = {1710.08424},
 primaryClass = {astro-ph.SR},
       adsurl = {https://ui.adsabs.harvard.edu/abs/2018ApJS..234...34P}
}

@ARTICLE{mesa2015,
       author = {{Paxton}, Bill and {Marchant}, Pablo and {Schwab}, Josiah and {Bauer}, Evan B. and {Bildsten}, Lars and {Cantiello}, Matteo and {Dessart}, Luc and {Farmer}, R. and {Hu}, H. and {Langer}, N. and {Townsend}, R.~H.~D. and {Townsley}, Dean M. and {Timmes}, F.~X.},
        title = "{Modules for Experiments in Stellar Astrophysics (MESA): Binaries, Pulsations, and Explosions}",
      journal = {\apjs},
         year = 2015,
        month = sep,
       volume = {220},
       number = {1},
          eid = {15},
        pages = {15},
          doi = {10.1088/0067-0049/220/1/15},
archivePrefix = {arXiv},
       eprint = {1506.03146},
 primaryClass = {astro-ph.SR},
       adsurl = {https://ui.adsabs.harvard.edu/abs/2015ApJS..220...15P}
}

@ARTICLE{mesa2013,
       author = {{Paxton}, Bill and {Cantiello}, Matteo and {Arras}, Phil and {Bildsten}, Lars and {Brown}, Edward F. and {Dotter}, Aaron and {Mankovich}, Christopher and {Montgomery}, M.~H. and {Stello}, Dennis and {Timmes}, F.~X. and {Townsend}, Richard},
        title = "{Modules for Experiments in Stellar Astrophysics (MESA): Planets, Oscillations, Rotation, and Massive Stars}",
      journal = {\apjs},
         year = 2013,
        month = sep,
       volume = {208},
       number = {1},
          eid = {4},
        pages = {4},
          doi = {10.1088/0067-0049/208/1/4},
archivePrefix = {arXiv},
       eprint = {1301.0319},
 primaryClass = {astro-ph.SR},
       adsurl = {https://ui.adsabs.harvard.edu/abs/2013ApJS..208....4P}
}

@ARTICLE{mesa2011,
       author = {{Paxton}, Bill and {Bildsten}, Lars and {Dotter}, Aaron and {Herwig}, Falk and {Lesaffre}, Pierre and {Timmes}, Frank},
        title = "{Modules for Experiments in Stellar Astrophysics (MESA)}",
      journal = {\apjs},
         year = 2011,
        month = jan,
       volume = {192},
       number = {1},
          eid = {3},
        pages = {3},
          doi = {10.1088/0067-0049/192/1/3},
archivePrefix = {arXiv},
       eprint = {1009.1622},
 primaryClass = {astro-ph.SR},
       adsurl = {https://ui.adsabs.harvard.edu/abs/2011ApJS..192....3P}
}

@ARTICLE{gyre2013,
       author = {{Townsend}, R.~H.~D. and {Teitler}, S.~A.},
        title = "{GYRE: an open-source stellar oscillation code based on a new Magnus Multiple Shooting scheme}",
      journal = {\mnras},
         year = 2013,
        month = nov,
       volume = {435},
       number = {4},
        pages = {3406-3418},
          doi = {10.1093/mnras/stt1533},
archivePrefix = {arXiv},
       eprint = {1308.2965},
 primaryClass = {astro-ph.SR},
       adsurl = {https://ui.adsabs.harvard.edu/abs/2013MNRAS.435.3406T}
}

@ARTICLE{BeddingFirstResultKepler,
       author = {{Bedding}, T.~R. and {Huber}, D. and {Stello}, D. and {Elsworth}, Y.~P. and {Hekker}, S. and {Kallinger}, T. and {Mathur}, S. and {Mosser}, B. and {Preston}, H.~L. and {Ballot}, J. and {Barban}, C. and {Broomhall}, A.~M. and {Buzasi}, D.~L. and {Chaplin}, W.~J. and {Garc{\'\i}a}, R.~A. and {Gruberbauer}, M. and {Hale}, S.~J. and {De Ridder}, J. and {Frandsen}, S. and {Borucki}, W.~J. and {Brown}, T. and {Christensen-Dalsgaard}, J. and {Gilliland}, R.~L. and {Jenkins}, J.~M. and {Kjeldsen}, H. and {Koch}, D. and {Belkacem}, K. and {Bildsten}, L. and {Bruntt}, H. and {Campante}, T.~L. and {Deheuvels}, S. and {Derekas}, A. and {Dupret}, M. -A. and {Goupil}, M. -J. and {Hatzes}, A. and {Houdek}, G. and {Ireland}, M.~J. and {Jiang}, C. and {Karoff}, C. and {Kiss}, L.~L. and {Lebreton}, Y. and {Miglio}, A. and {Montalb{\'a}n}, J. and {Noels}, A. and {Roxburgh}, I.~W. and {Sangaralingam}, V. and {Stevens}, I.~R. and {Suran}, M.~D. and {Tarrant}, N.~J. and {Weiss}, A.},
        title = "{Solar-like Oscillations in Low-luminosity Red Giants: First Results from Kepler}",
      journal = {\apjl},
         year = 2010,
        month = apr,
       volume = {713},
       number = {2},
        pages = {L176-L181},
          doi = {10.1088/2041-8205/713/2/L176},
archivePrefix = {arXiv},
       eprint = {1001.0229},
 primaryClass = {astro-ph.SR},
       adsurl = {https://ui.adsabs.harvard.edu/abs/2010ApJ...713L.176B}
}

@ARTICLE{Kallinger2012,
       author = {{Kallinger}, T. and {Hekker}, S. and {Mosser}, B. and {De Ridder}, J. and {Bedding}, T.~R. and {Elsworth}, Y.~P. and {Gruberbauer}, M. and {Guenther}, D.~B. and {Stello}, D. and {Basu}, S. and {Garc{\'\i}a}, R.~A. and {Chaplin}, W.~J. and {Mullally}, F. and {Still}, M. and {Thompson}, S.~E.},
        title = "{Evolutionary influences on the structure of red-giant acoustic oscillation spectra from 600d of Kepler observations}",
      journal = {\aap},
         year = 2012,
        month = may,
       volume = {541},
          eid = {A51},
        pages = {A51},
          doi = {10.1051/0004-6361/201218854},
archivePrefix = {arXiv},
       eprint = {1203.3134},
 primaryClass = {astro-ph.SR},
       adsurl = {https://ui.adsabs.harvard.edu/abs/2012A&A...541A..51K}
}

@ARTICLE{Sreenivas2024,
       author = {{Sreenivas}, K.~R. and {Bedding}, Timothy R. and {Li}, Yaguang and {Huber}, Daniel and {Crawford}, Courtney L. and {Stello}, Dennis and {Yu}, Jie},
        title = "{A simple method to measure {\ensuremath{\nu}}$_{max}$ for asteroseismology: application to 16 000 oscillating Kepler red giants}",
      journal = {\mnras},
         year = 2024,
        month = may,
       volume = {530},
       number = {3},
        pages = {3477-3487},
          doi = {10.1093/mnras/stae991},
archivePrefix = {arXiv},
       eprint = {2401.17557},
 primaryClass = {astro-ph.SR},
       adsurl = {https://ui.adsabs.harvard.edu/abs/2024MNRAS.530.3477S}
}

@ARTICLE{Yu2018,
       author = {{Yu}, Jie and {Huber}, Daniel and {Bedding}, Timothy R. and {Stello}, Dennis and {Hon}, Marc and {Murphy}, Simon J. and {Khanna}, Shourya},
        title = "{Asteroseismology of 16,000 Kepler Red Giants: Global Oscillation Parameters, Masses, and Radii}",
      journal = {\apjs},
         year = 2018,
        month = jun,
       volume = {236},
       number = {2},
          eid = {42},
        pages = {42},
          doi = {10.3847/1538-4365/aaaf74},
archivePrefix = {arXiv},
       eprint = {1802.04455},
 primaryClass = {astro-ph.SR},
       adsurl = {https://ui.adsabs.harvard.edu/abs/2018ApJS..236...42Y}
}

@ARTICLE{Montalban2010,
       author = {{Montalb{\'a}n}, J. and {Miglio}, A. and {Noels}, A. and {Scuflaire}, R. and {Ventura}, P.},
        title = "{Seismic Diagnostics of Red Giants: First Comparison with Stellar Models}",
      journal = {\apjl},
         year = 2010,
        month = oct,
       volume = {721},
       number = {2},
        pages = {L182-L188},
          doi = {10.1088/2041-8205/721/2/L182},
archivePrefix = {arXiv},
       eprint = {1009.1754},
 primaryClass = {astro-ph.SR},
       adsurl = {https://ui.adsabs.harvard.edu/abs/2010ApJ...721L.182M}
}

@INPROCEEDINGS{Borucki2008Kepler,
       author = {{Borucki}, William and {Koch}, David and {Basri}, Gibor and {Batalha}, Natalie and {Brown}, Timothy and {Caldwell}, Douglas and {Christensen-Dalsgaard}, J{\o}rgen and {Cochran}, William and {Dunham}, Edward and {Gautier}, Thomas N. and {Geary}, John and {Gilliland}, Ronald and {Jenkins}, Jon and {Kondo}, Yoji and {Latham}, David and {Lissauer}, Jack J. and {Monet}, David},
        title = "{Finding Earth-size planets in the habitable zone: the Kepler Mission}",
    booktitle = {Exoplanets: Detection, Formation and Dynamics},
         year = 2008,
       editor = {{Sun}, Yi-Sui and {Ferraz-Mello}, Sylvio and {Zhou}, Ji-Lin},
       series = {IAU Symposium},
       volume = {249},
        month = may,
        pages = {17-24},
          doi = {10.1017/S174392130801630X},
       adsurl = {https://ui.adsabs.harvard.edu/abs/2008IAUS..249...17B}
}

@ARTICLE{Handberg2017,
       author = {{Handberg}, R. and {Brogaard}, K. and {Miglio}, A. and {Bossini}, D. and {Elsworth}, Y. and {Slumstrup}, D. and {Davies}, G.~R. and {Chaplin}, W.~J.},
        title = "{NGC 6819: testing the asteroseismic mass scale, mass loss and evidence for products of non-standard evolution}",
      journal = {\mnras},
         year = 2017,
        month = nov,
       volume = {472},
       number = {1},
        pages = {979-997},
          doi = {10.1093/mnras/stx1929},
archivePrefix = {arXiv},
       eprint = {1707.08223},
 primaryClass = {astro-ph.SR},
       adsurl = {https://ui.adsabs.harvard.edu/abs/2017MNRAS.472..979H}
}

@ARTICLE{White2011,
       author = {{White}, Timothy R. and {Bedding}, Timothy R. and {Stello}, Dennis and {Christensen-Dalsgaard}, J{\o}rgen and {Huber}, Daniel and {Kjeldsen}, Hans},
        title = "{Calculating Asteroseismic Diagrams for Solar-like Oscillations}",
      journal = {\apj},
         year = 2011,
        month = dec,
       volume = {743},
       number = {2},
          eid = {161},
        pages = {161},
          doi = {10.1088/0004-637X/743/2/161},
archivePrefix = {arXiv},
       eprint = {1109.3455},
 primaryClass = {astro-ph.SR},
       adsurl = {https://ui.adsabs.harvard.edu/abs/2011ApJ...743..161W}
}

@ARTICLE{Grec1983,
       author = {{Grec}, G. and {Fossat}, E. and {Pomerantz}, M.~A.},
        title = "{Full-Disk Observations of Solar Oscillations from the Geographic South-Pole - Latest Results}",
      journal = {\solphys},
         year = 1983,
        month = jan,
       volume = {82},
       number = {1-2},
        pages = {55-66},
          doi = {10.1007/BF00145545},
       adsurl = {https://ui.adsabs.harvard.edu/abs/1983SoPh...82...55G}
}

@ARTICLE{Huber2011,
       author = {{Huber}, D. and {Bedding}, T.~R. and {Stello}, D. and {Hekker}, S. and {Mathur}, S. and {Mosser}, B. and {Verner}, G.~A. and {Bonanno}, A. and {Buzasi}, D.~L. and {Campante}, T.~L. and {Elsworth}, Y.~P. and {Hale}, S.~J. and {Kallinger}, T. and {Silva Aguirre}, V. and {Chaplin}, W.~J. and {De Ridder}, J. and {Garc{\'\i}a}, R.~A. and {Appourchaux}, T. and {Frandsen}, S. and {Houdek}, G. and {Molenda-{\.Z}akowicz}, J. and {Monteiro}, M.~J.~P.~F.~G. and {Christensen-Dalsgaard}, J. and {Gilliland}, R.~L. and {Kawaler}, S.~D. and {Kjeldsen}, H. and {Broomhall}, A.~M. and {Corsaro}, E. and {Salabert}, D. and {Sanderfer}, D.~T. and {Seader}, S.~E. and {Smith}, J.~C.},
        title = "{Testing Scaling Relations for Solar-like Oscillations from the Main Sequence to Red Giants Using Kepler Data}",
      journal = {\apj},
         year = 2011,
        month = dec,
       volume = {743},
       number = {2},
          eid = {143},
        pages = {143},
          doi = {10.1088/0004-637X/743/2/143},
archivePrefix = {arXiv},
       eprint = {1109.3460},
 primaryClass = {astro-ph.SR},
       adsurl = {https://ui.adsabs.harvard.edu/abs/2011ApJ...743..143H}
}

@ARTICLE{Huber2010,
       author = {{Huber}, D. and {Bedding}, T.~R. and {Stello}, D. and {Mosser}, B. and {Mathur}, S. and {Kallinger}, T. and {Hekker}, S. and {Elsworth}, Y.~P. and {Buzasi}, D.~L. and {De Ridder}, J. and {Gilliland}, R.~L. and {Kjeldsen}, H. and {Chaplin}, W.~J. and {Garc{\'\i}a}, R.~A. and {Hale}, S.~J. and {Preston}, H.~L. and {White}, T.~R. and {Borucki}, W.~J. and {Christensen-Dalsgaard}, J. and {Clarke}, B.~D. and {Jenkins}, J.~M. and {Koch}, D.},
        title = "{Asteroseismology of Red Giants from the First Four Months of Kepler Data: Global Oscillation Parameters for 800 Stars}",
      journal = {\apj},
         year = 2010,
        month = nov,
       volume = {723},
       number = {2},
        pages = {1607-1617},
          doi = {10.1088/0004-637X/723/2/1607},
archivePrefix = {arXiv},
       eprint = {1010.4566},
 primaryClass = {astro-ph.SR},
       adsurl = {https://ui.adsabs.harvard.edu/abs/2010ApJ...723.1607H}
}

@ARTICLE{Houdek1999,
       author = {{Houdek}, G. and {Balmforth}, N.~J. and {Christensen-Dalsgaard}, J. and {Gough}, D.~O.},
        title = "{Amplitudes of stochastically excited oscillations in main-sequence stars}",
      journal = {\aap},
         year = 1999,
        month = nov,
       volume = {351},
        pages = {582-596},
          doi = {10.48550/arXiv.astro-ph/9909107},
archivePrefix = {arXiv},
       eprint = {astro-ph/9909107},
 primaryClass = {astro-ph},
       adsurl = {https://ui.adsabs.harvard.edu/abs/1999A&A...351..582H}
}

@ARTICLE{Samadi_Goupil_2001,
       author = {{Samadi}, R. and {Goupil}, M. -J.},
        title = "{Excitation of stellar p-modes by turbulent convection. I. Theoretical formulation}",
      journal = {\aap},
         year = 2001,
        month = apr,
       volume = {370},
        pages = {136-146},
          doi = {10.1051/0004-6361:20010212},
archivePrefix = {arXiv},
       eprint = {astro-ph/0101109},
 primaryClass = {astro-ph},
       adsurl = {https://ui.adsabs.harvard.edu/abs/2001A&A...370..136S}
}

@ARTICLE{Bedding2011Natur.471..608B,
       author = {{Bedding}, Timothy R. and {Mosser}, Benoit and {Huber}, Daniel and {Montalb{\'a}n}, Josefina and {Beck}, Paul and {Christensen-Dalsgaard}, J{\o}rgen and {Elsworth}, Yvonne P. and {Garc{\'\i}a}, Rafael A. and {Miglio}, Andrea and {Stello}, Dennis and {White}, Timothy R. and {De Ridder}, Joris and {Hekker}, Saskia and {Aerts}, Conny and {Barban}, Caroline and {Belkacem}, Kevin and {Broomhall}, Anne-Marie and {Brown}, Timothy M. and {Buzasi}, Derek L. and {Carrier}, Fabien and {Chaplin}, William J. and {di Mauro}, Maria Pia and {Dupret}, Marc-Antoine and {Frandsen}, S{\o}ren and {Gilliland}, Ronald L. and {Goupil}, Marie-Jo and {Jenkins}, Jon M. and {Kallinger}, Thomas and {Kawaler}, Steven and {Kjeldsen}, Hans and {Mathur}, Savita and {Noels}, Arlette and {Silva Aguirre}, Victor and {Ventura}, Paolo},
        title = "{Gravity modes as a way to distinguish between hydrogen- and helium-burning red giant stars}",
      journal = {\nat},
         year = 2011,
        month = mar,
       volume = {471},
       number = {7340},
        pages = {608-611},
          doi = {10.1038/nature09935},
archivePrefix = {arXiv},
       eprint = {1103.5805},
 primaryClass = {astro-ph.SR},
       adsurl = {https://ui.adsabs.harvard.edu/abs/2011Natur.471..608B}
}

@ARTICLE{Hon2024ApJ...973..154H,
       author = {{Hon}, Marc and {Li}, Yaguang and {Ong}, Joel},
        title = "{Flow-based Generative Emulation of Grids of Stellar Evolutionary Models}",
      journal = {\apj},
         year = 2024,
        month = oct,
       volume = {973},
       number = {2},
          eid = {154},
        pages = {154},
          doi = {10.3847/1538-4357/ad6320},
archivePrefix = {arXiv},
       eprint = {2407.09427},
 primaryClass = {astro-ph.SR},
       adsurl = {https://ui.adsabs.harvard.edu/abs/2024ApJ...973..154H}
}

@ARTICLE{Pinsonneault2025,
       author = {{Pinsonneault}, Marc H. and {Zinn}, Joel C. and {Tayar}, Jamie and {Serenelli}, Aldo and {Garc{\'\i}a}, Rafael A. and {Mathur}, Savita and {Vrard}, Mathieu and {Elsworth}, Yvonne P. and {Mosser}, Benoit and {Stello}, Dennis and {Bell}, Keaton J. and {Bugnet}, Lisa and {Corsaro}, Enrico and {Gaulme}, Patrick and {Hekker}, Saskia and {Hon}, Marc and {Huber}, Daniel and {Kallinger}, Thomas and {Cao}, Kaili and {Johnson}, Jennifer A. and {Liagre}, Bastien and {Patton}, Rachel A. and {Santos}, {\^A}ngela R.~G. and {Basu}, Sarbani and {Beck}, Paul G. and {Beers}, Timothy C. and {Chaplin}, William J. and {Cunha}, Katia and {Frinchaboy}, Peter M. and {Girardi}, L{\'e}o and {Godoy-Rivera}, Diego and {Holtzman}, Jon A. and {J{\"o}nsson}, Henrik and {M{\'e}sz{\'a}ros}, Szabolcs and {Reyes}, Claudia and {Rix}, Hans-Walter and {Shetrone}, Matthew and {Smith}, Verne V. and {Spoo}, Taylor and {Stassun}, Keivan G. and {Wang}, Ji},
        title = "{APOKASC-3: The Third Joint Spectroscopic and Asteroseismic Catalog for Evolved Stars in the Kepler Fields}",
      journal = {\apjs},
         year = 2025,
        month = feb,
       volume = {276},
       number = {2},
          eid = {69},
        pages = {69},
          doi = {10.3847/1538-4365/ad9fef},
archivePrefix = {arXiv},
       eprint = {2410.00102},
 primaryClass = {astro-ph.SR},
       adsurl = {https://ui.adsabs.harvard.edu/abs/2025ApJS..276...69P}
}

@ARTICLE{Vrard2015A&A...579A..84V,
       author = {{Vrard}, M. and {Mosser}, B. and {Barban}, C. and {Belkacem}, K. and {Elsworth}, Y. and {Kallinger}, T. and {Hekker}, S. and {Samadi}, R. and {Beck}, P.~G.},
        title = "{Helium signature in red giant oscillation patterns observed by Kepler}",
      journal = {\aap},
         year = 2015,
        month = jul,
       volume = {579},
          eid = {A84},
        pages = {A84},
          doi = {10.1051/0004-6361/201425064},
archivePrefix = {arXiv},
       eprint = {1505.07280},
 primaryClass = {astro-ph.SR},
       adsurl = {https://ui.adsabs.harvard.edu/abs/2015A&A...579A..84V}
}

@ARTICLE{Corsaro2012ApJ...757..190C,
       author = {{Corsaro}, Enrico and {Stello}, Dennis and {Huber}, Daniel and {Bedding}, Timothy R. and {Bonanno}, Alfio and {Brogaard}, Karsten and {Kallinger}, Thomas and {Benomar}, Othman and {White}, Timothy R. and {Mosser}, Benoit and {Basu}, Sarbani and {Chaplin}, William J. and {Christensen-Dalsgaard}, J{\o}rgen and {Elsworth}, Yvonne P. and {Garc{\'\i}a}, Rafael A. and {Hekker}, Saskia and {Kjeldsen}, Hans and {Mathur}, Savita and {Meibom}, S{\o}ren and {Hall}, Jennifer R. and {Ibrahim}, Khadeejah A. and {Klaus}, Todd C.},
        title = "{Asteroseismology of the Open Clusters NGC 6791, NGC 6811, and NGC 6819 from 19 Months of Kepler Photometry}",
      journal = {\apj},
         year = 2012,
        month = oct,
       volume = {757},
       number = {2},
          eid = {190},
        pages = {190},
          doi = {10.1088/0004-637X/757/2/190},
archivePrefix = {arXiv},
       eprint = {1205.4023},
 primaryClass = {astro-ph.SR},
       adsurl = {https://ui.adsabs.harvard.edu/abs/2012ApJ...757..190C}
}

@ARTICLE{Roxburgh2003A&A...411..215R,
       author = {{Roxburgh}, I.~W. and {Vorontsov}, S.~V.},
        title = "{The ratio of small to large separations of acoustic oscillations as a diagnostic of the interior of solar-like stars}",
      journal = {\aap},
         year = 2003,
        month = nov,
       volume = {411},
        pages = {215-220},
          doi = {10.1051/0004-6361:20031318},
       adsurl = {https://ui.adsabs.harvard.edu/abs/2003A&A...411..215R}
}

@ARTICLE{Roxburgh2005A&A...434..665R,
       author = {{Roxburgh}, I.~W.},
        title = "{The ratio of small to large separations of stellar p-modes}",
      journal = {\aap},
         year = 2005,
        month = may,
       volume = {434},
       number = {2},
        pages = {665-669},
          doi = {10.1051/0004-6361:20041957},
       adsurl = {https://ui.adsabs.harvard.edu/abs/2005A&A...434..665R}
}

@ARTICLE{Floranes2005MNRAS.356..671O,
       author = {{Ot{\'\i} Floranes}, H. and {Christensen-Dalsgaard}, J. and {Thompson}, M.~J.},
        title = "{The use of frequency-separation ratios for asteroseismology}",
      journal = {\mnras},
         year = 2005,
        month = jan,
       volume = {356},
       number = {2},
        pages = {671-679},
          doi = {10.1111/j.1365-2966.2004.08487.x},
       adsurl = {https://ui.adsabs.harvard.edu/abs/2005MNRAS.356..671O}
}

@INPROCEEDINGS{Baglin2006cosp...36.3749B,
       author = {{Baglin}, A. and {Auvergne}, M. and {Boisnard}, L. and {Lam-Trong}, T. and {Barge}, P. and {Catala}, C. and {Deleuil}, M. and {Michel}, E. and {Weiss}, W.},
        title = "{CoRoT: a high precision photometer for stellar ecolution and exoplanet finding}",
    booktitle = {36th COSPAR Scientific Assembly},
         year = 2006,
       volume = {36},
        month = jan,
        pages = {3749},
       adsurl = {https://ui.adsabs.harvard.edu/abs/2006cosp...36.3749B}
}

@ARTICLE{Ricker2015JATIS...1a4003R,
       author = {{Ricker}, George R. and {Winn}, Joshua N. and {Vanderspek}, Roland and {Latham}, David W. and {Bakos}, G{\'a}sp{\'a}r {\'A}. and {Bean}, Jacob L. and {Berta-Thompson}, Zachory K. and {Brown}, Timothy M. and {Buchhave}, Lars and {Butler}, Nathaniel R. and {Butler}, R. Paul and {Chaplin}, William J. and {Charbonneau}, David and {Christensen-Dalsgaard}, J{\o}rgen and {Clampin}, Mark and {Deming}, Drake and {Doty}, John and {De Lee}, Nathan and {Dressing}, Courtney and {Dunham}, Edward W. and {Endl}, Michael and {Fressin}, Francois and {Ge}, Jian and {Henning}, Thomas and {Holman}, Matthew J. and {Howard}, Andrew W. and {Ida}, Shigeru and {Jenkins}, Jon M. and {Jernigan}, Garrett and {Johnson}, John Asher and {Kaltenegger}, Lisa and {Kawai}, Nobuyuki and {Kjeldsen}, Hans and {Laughlin}, Gregory and {Levine}, Alan M. and {Lin}, Douglas and {Lissauer}, Jack J. and {MacQueen}, Phillip and {Marcy}, Geoffrey and {McCullough}, Peter R. and {Morton}, Timothy D. and {Narita}, Norio and {Paegert}, Martin and {Palle}, Enric and {Pepe}, Francesco and {Pepper}, Joshua and {Quirrenbach}, Andreas and {Rinehart}, Stephen A. and {Sasselov}, Dimitar and {Sato}, Bun'ei and {Seager}, Sara and {Sozzetti}, Alessandro and {Stassun}, Keivan G. and {Sullivan}, Peter and {Szentgyorgyi}, Andrew and {Torres}, Guillermo and {Udry}, Stephane and {Villasenor}, Joel},
        title = "{Transiting Exoplanet Survey Satellite (TESS)}",
      journal = {Journal of Astronomical Telescopes, Instruments, and Systems},
         year = 2015,
        month = jan,
       volume = {1},
          eid = {014003},
        pages = {014003},
          doi = {10.1117/1.JATIS.1.1.014003},
       adsurl = {https://ui.adsabs.harvard.edu/abs/2015JATIS...1a4003R}
}

@ARTICLE{Chaplin2013ARA&A..51..353C,
       author = {{Chaplin}, William J. and {Miglio}, Andrea},
        title = "{Asteroseismology of Solar-Type and Red-Giant Stars}",
      journal = {\araa},
         year = 2013,
        month = aug,
       volume = {51},
       number = {1},
        pages = {353-392},
          doi = {10.1146/annurev-astro-082812-140938},
archivePrefix = {arXiv},
       eprint = {1303.1957},
 primaryClass = {astro-ph.SR},
       adsurl = {https://ui.adsabs.harvard.edu/abs/2013ARA&A..51..353C}
}

@ARTICLE{Tassoul1980ApJS...43..469T,
       author = {{Tassoul}, M.},
        title = "{Asymptotic approximations for stellar nonradial pulsations.}",
      journal = {\apjs},
         year = 1980,
        month = aug,
       volume = {43},
        pages = {469-490},
          doi = {10.1086/190678},
       adsurl = {https://ui.adsabs.harvard.edu/abs/1980ApJS...43..469T}
}

@ARTICLE{Ong2025ApJ...980..199O,
       author = {{Ong}, J.~M. Joel and {Lindsay}, Christopher J. and {Reyes}, Claudia and {Stello}, Dennis and {Roxburgh}, Ian W.},
        title = "{Resolving an Asteroseismic Catastrophe: Structural Diagnostics from p-mode Phase Functions off the Main Sequence}",
      journal = {\apj},
         year = 2025,
        month = feb,
       volume = {980},
       number = {2},
          eid = {199},
        pages = {199},
          doi = {10.3847/1538-4357/ada949},
archivePrefix = {arXiv},
       eprint = {2501.05343},
 primaryClass = {astro-ph.SR},
       adsurl = {https://ui.adsabs.harvard.edu/abs/2025ApJ...980..199O}
}

@ARTICLE{Stello2016Natur.529..364S,
       author = {{Stello}, Dennis and {Cantiello}, Matteo and {Fuller}, Jim and {Huber}, Daniel and {Garc{\'\i}a}, Rafael A. and {Bedding}, Timothy R. and {Bildsten}, Lars and {Silva Aguirre}, Victor},
        title = "{A prevalence of dynamo-generated magnetic fields in the cores of intermediate-mass stars}",
      journal = {\nat},
         year = 2016,
        month = jan,
       volume = {529},
       number = {7586},
        pages = {364-367},
          doi = {10.1038/nature16171},
archivePrefix = {arXiv},
       eprint = {1601.00004},
 primaryClass = {astro-ph.SR},
       adsurl = {https://ui.adsabs.harvard.edu/abs/2016Natur.529..364S}
}

@INPROCEEDINGS{Christensen-Dalsgaard1988IAUS..123..295C,
       author = {{Christensen-Dalsgaard}, J.},
        title = "{A Hertzsprung-Russell Diagram for Stellar Oscillations}",
    booktitle = {Advances in Helio- and Asteroseismology},
         year = 1988,
       editor = {{Christensen-Dalsgaard}, Jorgen and {Frandsen}, Soren},
       series = {IAU Symposium},
       volume = {123},
        month = jan,
        pages = {295},
       adsurl = {https://ui.adsabs.harvard.edu/abs/1988IAUS..123..295C}
}

@ARTICLE{Girardi2016ARA&A..54...95G,
       author = {{Girardi}, L{\'e}o},
        title = "{Red Clump Stars}",
      journal = {\araa},
         year = 2016,
        month = sep,
       volume = {54},
        pages = {95-133},
          doi = {10.1146/annurev-astro-081915-023354},
       adsurl = {https://ui.adsabs.harvard.edu/abs/2016ARA&A..54...95G}
}

@ARTICLE{Smith2012PASP..124.1000S,
       author = {{Smith}, Jeffrey C. and {Stumpe}, Martin C. and {Van Cleve}, Jeffrey E. and {Jenkins}, Jon M. and {Barclay}, Thomas S. and {Fanelli}, Michael N. and {Girouard}, Forrest R. and {Kolodziejczak}, Jeffery J. and {McCauliff}, Sean D. and {Morris}, Robert L. and {Twicken}, Joseph D.},
        title = "{Kepler Presearch Data Conditioning II - A Bayesian Approach to Systematic Error Correction}",
      journal = {\pasp},
         year = 2012,
        month = sep,
       volume = {124},
       number = {919},
        pages = {1000},
          doi = {10.1086/667697},
archivePrefix = {arXiv},
       eprint = {1203.1383},
 primaryClass = {astro-ph.IM},
       adsurl = {https://ui.adsabs.harvard.edu/abs/2012PASP..124.1000S}
}

@ARTICLE{Stumpe2012PASP..124..985S,
       author = {{Stumpe}, Martin C. and {Smith}, Jeffrey C. and {Van Cleve}, Jeffrey E. and {Twicken}, Joseph D. and {Barclay}, Thomas S. and {Fanelli}, Michael N. and {Girouard}, Forrest R. and {Jenkins}, Jon M. and {Kolodziejczak}, Jeffery J. and {McCauliff}, Sean D. and {Morris}, Robert L.},
        title = "{Kepler Presearch Data Conditioning I{\textemdash}Architecture and Algorithms for Error Correction in Kepler Light Curves}",
      journal = {\pasp},
         year = 2012,
        month = sep,
       volume = {124},
       number = {919},
        pages = {985},
          doi = {10.1086/667698},
archivePrefix = {arXiv},
       eprint = {1203.1382},
 primaryClass = {astro-ph.IM},
       adsurl = {https://ui.adsabs.harvard.edu/abs/2012PASP..124..985S}
}

@ARTICLE{Stello2016PASA...33...11S,
       author = {{Stello}, Dennis and {Cantiello}, Matteo and {Fuller}, Jim and {Garcia}, Rafael A. and {Huber}, Daniel},
        title = "{Suppression of Quadrupole and Octupole Modes in Red Giants Observed by Kepler *}",
      journal = {\pasa},
         year = 2016,
        month = mar,
       volume = {33},
          eid = {e011},
        pages = {e011},
          doi = {10.1017/pasa.2016.9},
archivePrefix = {arXiv},
       eprint = {1602.05193},
 primaryClass = {astro-ph.SR},
       adsurl = {https://ui.adsabs.harvard.edu/abs/2016PASA...33...11S}
}

@ARTICLE{Li2023MNRAS.523..916L,
       author = {{Li}, Yaguang and {Bedding}, Timothy R. and {Stello}, Dennis and {Huber}, Daniel and {Hon}, Marc and {Joyce}, Meridith and {Li}, Tanda and {Perkins}, Jean and {White}, Timothy R. and {Zinn}, Joel C. and {Howard}, Andrew W. and {Isaacson}, Howard and {Hey}, Daniel R. and {Kjeldsen}, Hans},
        title = "{A prescription for the asteroseismic surface correction}",
      journal = {\mnras},
         year = 2023,
        month = jul,
       volume = {523},
       number = {1},
        pages = {916-927},
          doi = {10.1093/mnras/stad1445},
archivePrefix = {arXiv},
       eprint = {2208.01176},
 primaryClass = {astro-ph.SR},
       adsurl = {https://ui.adsabs.harvard.edu/abs/2023MNRAS.523..916L}
}

@ARTICLE{Ball2014A&A...568A.123B,
       author = {{Ball}, W.~H. and {Gizon}, L.},
        title = "{A new correction of stellar oscillation frequencies for near-surface effects}",
      journal = {\aap},
         year = 2014,
        month = aug,
       volume = {568},
          eid = {A123},
        pages = {A123},
          doi = {10.1051/0004-6361/201424325},
archivePrefix = {arXiv},
       eprint = {1408.0986},
 primaryClass = {astro-ph.SR},
       adsurl = {https://ui.adsabs.harvard.edu/abs/2014A&A...568A.123B}
}

@ARTICLE{Ball2017A&A...600A.128B,
       author = {{Ball}, W.~H. and {Gizon}, L.},
        title = "{Surface-effect corrections for oscillation frequencies of evolved stars}",
      journal = {\aap},
         year = 2017,
        month = apr,
       volume = {600},
          eid = {A128},
        pages = {A128},
          doi = {10.1051/0004-6361/201630260},
archivePrefix = {arXiv},
       eprint = {1702.02570},
 primaryClass = {astro-ph.SR},
       adsurl = {https://ui.adsabs.harvard.edu/abs/2017A&A...600A.128B}
}

@ARTICLE{Dhanpal2023ApJ...958...63D,
       author = {{Dhanpal}, Siddharth and {Benomar}, Othman and {Hanasoge}, Shravan and {Takata}, Masao and {Panda}, Subrata Kumar and {Kundu}, Abhisek},
        title = "{Inferring Coupling Strengths of Mixed-mode Oscillations in Red Giant Stars Using Deep Learning}",
      journal = {\apj},
         year = 2023,
        month = nov,
       volume = {958},
       number = {1},
          eid = {63},
        pages = {63},
          doi = {10.3847/1538-4357/ad0046},
archivePrefix = {arXiv},
       eprint = {2309.17372},
 primaryClass = {astro-ph.SR},
       adsurl = {https://ui.adsabs.harvard.edu/abs/2023ApJ...958...63D}
}

@ARTICLE{Hon2019MNRAS.485.5616H,
       author = {{Hon}, Marc and {Stello}, Dennis and {Garc{\'\i}a}, Rafael A. and {Mathur}, Savita and {Sharma}, Sanjib and {Colman}, Isabel L. and {Bugnet}, Lisa},
        title = "{A search for red giant solar-like oscillations in all Kepler data}",
      journal = {\mnras},
         year = 2019,
        month = jun,
       volume = {485},
       number = {4},
        pages = {5616-5630},
          doi = {10.1093/mnras/stz622},
archivePrefix = {arXiv},
       eprint = {1903.00115},
 primaryClass = {astro-ph.SR},
       adsurl = {https://ui.adsabs.harvard.edu/abs/2019MNRAS.485.5616H}
}

@ARTICLE{Vrard2025A&A...697A.165V,
       author = {{Vrard}, M. and {Pinsonneault}, M.~H. and {Elsworth}, Y. and {Hon}, M. and {Kallinger}, T. and {Kuszlewicz}, J. and {Mosser}, B. and {Garc{\'\i}a}, R.~A. and {Tayar}, J. and {Bennett}, R. and {Cao}, K. and {Hekker}, S. and {Loyer}, L. and {Mathur}, S. and {Stello}, D.},
        title = "{Red giant evolutionary status determination: The complete Kepler catalog}",
      journal = {\aap},
         year = 2025,
        month = may,
       volume = {697},
          eid = {A165},
        pages = {A165},
          doi = {10.1051/0004-6361/202452635},
archivePrefix = {arXiv},
       eprint = {2411.03101},
 primaryClass = {astro-ph.SR},
       adsurl = {https://ui.adsabs.harvard.edu/abs/2025A&A...697A.165V}
}

@ARTICLE{Jackiewicz2021FrASS...7..102J,
       author = {{Jackiewicz}, Jason},
        title = "{Solar-Like Oscillators in the Kepler Era: A Review}",
      journal = {Frontiers in Astronomy and Space Sciences},
         year = 2021,
        month = mar,
       volume = {7},
          eid = {102},
        pages = {102},
          doi = {10.3389/fspas.2020.595017},
       adsurl = {https://ui.adsabs.harvard.edu/abs/2021FrASS...7..102J}
}

@INPROCEEDINGS{Bedding2012ASPC..462..195B,
       author = {{Bedding}, T.~R.},
        title = "{Replicated {\'E}chelle Diagrams in Asteroseismology: A Tool for Studying Mixed Modes and Avoided Crossings}",
    booktitle = {Progress in Solar/Stellar Physics with Helio- and Asteroseismology},
         year = 2012,
       editor = {{Shibahashi}, H. and {Takata}, M. and {Lynas-Gray}, A.~E.},
       series = {Astronomical Society of the Pacific Conference Series},
       volume = {462},
        month = sep,
        pages = {195},
          doi = {10.48550/arXiv.1109.5768},
archivePrefix = {arXiv},
       eprint = {1109.5768},
 primaryClass = {astro-ph.SR},
       adsurl = {https://ui.adsabs.harvard.edu/abs/2012ASPC..462..195B}
}

@ARTICLE{Reyes2025Natur.640..338R,
       author = {{Reyes}, Claudia and {Stello}, Dennis and {Ong}, Joel and {Lindsay}, Christopher and {Hon}, Marc and {Bedding}, Timothy R.},
        title = "{Acoustic modes in M67 cluster stars trace deepening convective envelopes}",
      journal = {\nat},
         year = 2025,
        month = apr,
       volume = {640},
       number = {8058},
        pages = {338-342},
          doi = {10.1038/s41586-025-08760-2},
archivePrefix = {arXiv},
       eprint = {2504.01828},
 primaryClass = {astro-ph.SR},
       adsurl = {https://ui.adsabs.harvard.edu/abs/2025Natur.640..338R}
}

@ARTICLE{Mosser2010A&A...517A..22M,
       author = {{Mosser}, B. and {Belkacem}, K. and {Goupil}, M. -J. and {Miglio}, A. and {Morel}, T. and {Barban}, C. and {Baudin}, F. and {Hekker}, S. and {Samadi}, R. and {De Ridder}, J. and {Weiss}, W. and {Auvergne}, M. and {Baglin}, A.},
        title = "{Red-giant seismic properties analyzed with CoRoT}",
      journal = {\aap},
         year = 2010,
        month = jul,
       volume = {517},
          eid = {A22},
        pages = {A22},
          doi = {10.1051/0004-6361/201014036},
archivePrefix = {arXiv},
       eprint = {1004.0449},
 primaryClass = {astro-ph.SR},
       adsurl = {https://ui.adsabs.harvard.edu/abs/2010A&A...517A..22M}
}

@inproceedings{Christensen-Dalsgaard2014aste.book..194C,
       author = {{Christensen-Dalsgaard}, J{\o}rgen},
        title = "{Asteroseismology of red giants}",
    booktitle = {Asteroseismology},
         year = 2014,
       editor = {{Pall{\'e}}, Pere L. and {Esteban}, Cesar},
        pages = {194},
          doi = {10.48550/arXiv.1106.5946},
       adsurl = {https://ui.adsabs.harvard.edu/abs/2014aste.book..194C}
}

@ARTICLE{Mosser2012A&A...537A..30M,
       author = {{Mosser}, B. and {Elsworth}, Y. and {Hekker}, S. and {Huber}, D. and {Kallinger}, T. and {Mathur}, S. and {Belkacem}, K. and {Goupil}, M.~J. and {Samadi}, R. and {Barban}, C. and {Bedding}, T.~R. and {Chaplin}, W.~J. and {Garc{\'\i}a}, R.~A. and {Stello}, D. and {De Ridder}, J. and {Middour}, C.~K. and {Morris}, R.~L. and {Quintana}, E.~V.},
        title = "{Characterization of the power excess of solar-like oscillations in red giants with Kepler}",
      journal = {\aap},
         year = 2012,
        month = jan,
       volume = {537},
          eid = {A30},
        pages = {A30},
          doi = {10.1051/0004-6361/20111735210.1086/141952},
archivePrefix = {arXiv},
       eprint = {1110.0980},
 primaryClass = {astro-ph.SR},
       adsurl = {https://ui.adsabs.harvard.edu/abs/2012A&A...537A..30M}
}

@ARTICLE{Christensen-Dalsgaard2014MNRAS.445.3685C,
       author = {{Christensen-Dalsgaard}, J{\o}rgen and {Silva Aguirre}, Victor and {Elsworth}, Yvonne and {Hekker}, Saskia},
        title = "{On the asymptotic acoustic-mode phase in red giant stars and its dependence on evolutionary state}",
      journal = {\mnras},
         year = 2014,
        month = dec,
       volume = {445},
       number = {4},
        pages = {3685-3693},
          doi = {10.1093/mnras/stu2007},
archivePrefix = {arXiv},
       eprint = {1409.7949},
 primaryClass = {astro-ph.SR},
       adsurl = {https://ui.adsabs.harvard.edu/abs/2014MNRAS.445.3685C}
}

@ARTICLE{OngBasu2019ApJ...885...26O,
       author = {{Ong}, J.~M. Joel and {Basu}, Sarbani},
        title = "{Structural and Evolutionary Diagnostics from Asteroseismic Phase Functions}",
      journal = {\apj},
         year = 2019,
        month = nov,
       volume = {885},
       number = {1},
          eid = {26},
        pages = {26},
          doi = {10.3847/1538-4357/ab425f},
archivePrefix = {arXiv},
       eprint = {1909.02580},
 primaryClass = {astro-ph.SR},
       adsurl = {https://ui.adsabs.harvard.edu/abs/2019ApJ...885...26O}
}

@ARTICLE{Girardi1999MNRAS.308..818G,
       author = {{Girardi}, L{\'e}o},
        title = "{A secondary clump of red giant stars: why and where}",
      journal = {\mnras},
         year = 1999,
        month = sep,
       volume = {308},
       number = {3},
        pages = {818-832},
          doi = {10.1046/j.1365-8711.1999.02746.x},
archivePrefix = {arXiv},
       eprint = {astro-ph/9901319},
 primaryClass = {astro-ph},
       adsurl = {https://ui.adsabs.harvard.edu/abs/1999MNRAS.308..818G}
}

@ARTICLE{Girardi2000A&AS..141..371G,
       author = {{Girardi}, L. and {Bressan}, A. and {Bertelli}, G. and {Chiosi}, C.},
        title = "{Evolutionary tracks and isochrones for low- and intermediate-mass stars: From 0.15 to 7 M$_{sun}$, and from Z=0.0004 to 0.03}",
      journal = {\aaps},
         year = 2000,
        month = feb,
       volume = {141},
        pages = {371-383},
          doi = {10.1051/aas:2000126},
archivePrefix = {arXiv},
       eprint = {astro-ph/9910164},
 primaryClass = {astro-ph},
       adsurl = {https://ui.adsabs.harvard.edu/abs/2000A&AS..141..371G}
}

@ARTICLE{Bressan2012MNRAS.427..127B,
       author = {{Bressan}, Alessandro and {Marigo}, Paola and {Girardi}, L{\'e}o. and {Salasnich}, Bernardo and {Dal Cero}, Claudia and {Rubele}, Stefano and {Nanni}, Ambra},
        title = "{PARSEC: stellar tracks and isochrones with the PAdova and TRieste Stellar Evolution Code}",
      journal = {\mnras},
         year = 2012,
        month = nov,
       volume = {427},
       number = {1},
        pages = {127-145},
          doi = {10.1111/j.1365-2966.2012.21948.x},
archivePrefix = {arXiv},
       eprint = {1208.4498},
 primaryClass = {astro-ph.SR},
       adsurl = {https://ui.adsabs.harvard.edu/abs/2012MNRAS.427..127B}
}

@ARTICLE{Constantino2015MNRAS.452..123C,
       author = {{Constantino}, Thomas and {Campbell}, Simon W. and {Christensen-Dalsgaard}, J{\o}rgen and {Lattanzio}, John C. and {Stello}, Dennis},
        title = "{The treatment of mixing in core helium burning models - I. Implications for asteroseismology}",
      journal = {\mnras},
         year = 2015,
        month = sep,
       volume = {452},
       number = {1},
        pages = {123-145},
          doi = {10.1093/mnras/stv1264},
archivePrefix = {arXiv},
       eprint = {1506.01209},
 primaryClass = {astro-ph.SR},
       adsurl = {https://ui.adsabs.harvard.edu/abs/2015MNRAS.452..123C}
}

@ARTICLE{Bossini2015MNRAS.453.2290B,
       author = {{Bossini}, Diego and {Miglio}, Andrea and {Salaris}, Maurizio and {Pietrinferni}, Adriano and {Montalb{\'a}n}, Josefina and {Bressan}, Alessandro and {Noels}, Arlette and {Cassisi}, Santi and {Girardi}, L{\'e}o and {Marigo}, Paola},
        title = "{Uncertainties on near-core mixing in red-clump stars: effects on the period spacing and on the luminosity of the AGB bump}",
      journal = {\mnras},
         year = 2015,
        month = nov,
       volume = {453},
       number = {3},
        pages = {2290-2301},
          doi = {10.1093/mnras/stv1738},
archivePrefix = {arXiv},
       eprint = {1507.07797},
 primaryClass = {astro-ph.SR},
       adsurl = {https://ui.adsabs.harvard.edu/abs/2015MNRAS.453.2290B}
}

@ARTICLE{Hekker2017A&ARv..25....1H,
       author = {{Hekker}, S. and {Christensen-Dalsgaard}, J.},
        title = "{Giant star seismology}",
      journal = {\aapr},
         year = 2017,
        month = jun,
       volume = {25},
       number = {1},
          eid = {1},
        pages = {1},
          doi = {10.1007/s00159-017-0101-x},
archivePrefix = {arXiv},
       eprint = {1609.07487},
 primaryClass = {astro-ph.SR},
       adsurl = {https://ui.adsabs.harvard.edu/abs/2017A&ARv..25....1H}
}

@ARTICLE{Zinn2019,
       author = {{Zinn}, Joel C. and {Stello}, Dennis and {Huber}, Daniel and {Sharma}, Sanjib},
        title = "{The Bayesian Asteroseismology Data Modeling Pipeline and Its Application to K2 Data}",
      journal = {\apj},
         year = 2019,
        month = oct,
       volume = {884},
       number = {2},
          eid = {107},
        pages = {107},
          doi = {10.3847/1538-4357/ab43c0},
archivePrefix = {arXiv},
       eprint = {1909.11927},
 primaryClass = {astro-ph.SR},
       adsurl = {https://ui.adsabs.harvard.edu/abs/2019ApJ...884..107Z}
}

@ARTICLE{Kallinger2019arXiv190609428K,
       author = {{Kallinger}, T.},
        title = "{Release note: Massive peak bagging of red giants in the Kepler field}",
      journal = {arXiv e-prints},
         year = 2019,
        month = jun,
          eid = {arXiv:1906.09428},
        pages = {arXiv:1906.09428},
          doi = {10.48550/arXiv.1906.09428},
archivePrefix = {arXiv},
       eprint = {1906.09428},
 primaryClass = {astro-ph.SR},
       adsurl = {https://ui.adsabs.harvard.edu/abs/2019arXiv190609428K}
}

@ARTICLE{Bouchy2002A&A...390..205B,
       author = {{Bouchy}, F. and {Carrier}, F.},
        title = "{The acoustic spectrum of alpha Cen A}",
      journal = {\aap},
         year = 2002,
        month = jul,
       volume = {390},
        pages = {205-212},
          doi = {10.1051/0004-6361:20020706},
archivePrefix = {arXiv},
       eprint = {astro-ph/0206051},
 primaryClass = {astro-ph},
       adsurl = {https://ui.adsabs.harvard.edu/abs/2002A&A...390..205B}
}

@ARTICLE{Bedding2004ApJ...614..380B,
       author = {{Bedding}, Timothy R. and {Kjeldsen}, Hans and {Butler}, R. Paul and {McCarthy}, Chris and {Marcy}, Geoffrey W. and {O'Toole}, Simon J. and {Tinney}, Christopher G. and {Wright}, Jason T.},
        title = "{Oscillation Frequencies and Mode Lifetimes in {\ensuremath{\alpha}} Centauri A}",
      journal = {\apj},
         year = 2004,
        month = oct,
       volume = {614},
       number = {1},
        pages = {380-385},
          doi = {10.1086/423484},
archivePrefix = {arXiv},
       eprint = {astro-ph/0406471},
 primaryClass = {astro-ph},
       adsurl = {https://ui.adsabs.harvard.edu/abs/2004ApJ...614..380B}
}

@misc{newville_2025_16175987,
  author       = {Newville, Matthew and
                  Otten, Renee and
                  Nelson, Andrew and
                  Stensitzki, Till and
                  Ingargiola, Antonino and
                  Allan, Daniel and
                  Fox, Austin and
                  Carter, Faustin and
                  Rawlik, Michal},
  title        = {LMFIT: Non-Linear Least-Squares Minimization and
                  Curve-Fitting for Python},
  year         = {2025},
  month        = jul,
  publisher    = {Zenodo},
  note         = {Version 1.3.4},
  doi          = {10.5281/zenodo.16175987},
  howpublished = {\url{https://doi.org/10.5281/zenodo.16175987}}
}

@article{Montalban2010ApJL721L182,
  author  = {Montalb{\'a}n, J. and Miglio, A. and Noels, A. and Scuflaire, R. and Ventura, P.},
  title   = {Seismic Diagnostics of Red Giants: First Comparison with Stellar Models},
  journal = {The Astrophysical Journal Letters},
  volume  = {721},
  number  = {2},
  pages   = {L182--L186},
  year    = {2010},
  doi     = {10.1088/2041-8205/721/2/L182}
}

@incollection{Montalban2012ASSP,
  author    = {Montalb{\'a}n, J. and Miglio, A. and Noels, A. and Scuflaire, R. and Ventura, P. and D'Antona, F.},
  title     = {Adiabatic Solar-Like Oscillations in Red Giant Stars},
  booktitle = {Red Giants as Probes of the Structure and Evolution of the Milky Way},
  editor    = {Miglio, A. and Montalb{\'a}n, J. and Noels, A.},
  series    = {Astrophysics and Space Science Proceedings},
  publisher = {Springer Berlin Heidelberg},
  address   = {Berlin, Heidelberg},
  year      = {2012},
  pages     = {23--32},
  doi       = {10.1007/978-3-642-18418-5_3}
}

@article{Miglio2021AA645A85,
  author  = {Miglio, A. and Chiappini, C. and Mackereth, T. and Brogaard, K. and Casagrande, L. and Chaplin, W. J. and Girardi, L. and Davies, G. R. and Kawata, D. and Khan, S. and Izzard, R. G. and Montalb{\'a}n, J. and Mosser, B. and Vincenzo, F. and Bossini, D. and Noels, A. and Rodrigues, T. S. and Valentini, M. and Mandel, I.},
  title   = {Age dissection of the Milky Way discs: {R}ed giants in the \emph{Kepler} field},
  journal = {Astronomy \& Astrophysics},
  volume  = {645},
  pages   = {A85},
  year    = {2021},
  doi     = {10.1051/0004-6361/202038307}
}

@article{Zhou2025ApJS27937,
  author  = {Zhou, Jianzhao and Bi, Shaolan and Li, Yaguang and Yu, Jie and Li, Tanda and Zhang, Xianfei and Ye, Lifei and Li, Mengjie and Liu, Long and Sun, Tiancheng and Chen, Yuqin},
  title   = {Asteroseismology of 687 TESS Red Giants: Individual Frequencies and Asymptotic Parameters},
  journal = {The Astrophysical Journal Supplement Series},
  year    = {2025},
  volume  = {279},
  pages   = {37},
  doi     = {10.3847/1538-4365/adde57}
}

@ARTICLE{Ball2018MNRAS.478.4697B,
       author = {{Ball}, W.~H. and {Theme{\ss}l}, N. and {Hekker}, S.},
        title = "{Surface effects on the red giant branch}",
      journal = {\mnras},
         year = 2018,
        month = aug,
       volume = {478},
       number = {4},
        pages = {4697-4709},
          doi = {10.1093/mnras/sty1141},
archivePrefix = {arXiv},
       eprint = {1804.11153},
 primaryClass = {astro-ph.SR},
       adsurl = {https://ui.adsabs.harvard.edu/abs/2018MNRAS.478.4697B}
}

@ARTICLE{Schimak2026MNRAS.546ag151S,
       author = {{Schimak}, Lea S. and {Bedding}, Timothy R. and {Crawford}, Courtney L. and {Beck}, Paul G. and {Li}, Yaguang and {Huber}, Daniel and {Ong}, Joel and {Montet}, Benjamin T. and {Pedersen}, May Gade and {Grossmann}, Desmond H. and {Mathur}, Savita and {Garc{\'\i}a}, Rafael A.},
        title = "{Testing red clump models with the asteroseismic binary KIC 10841730}",
      journal = {\mnras},
         year = 2026,
        month = mar,
       volume = {546},
       number = {3},
          eid = {stag151},
        pages = {stag151},
          doi = {10.1093/mnras/stag151},
archivePrefix = {arXiv},
       eprint = {2601.12773},
 primaryClass = {astro-ph.SR},
       adsurl = {https://ui.adsabs.harvard.edu/abs/2026MNRAS.546ag151S}
}

%\begin{thebibliography}{99}
%\end{thebibliography}

%%%%%%%%%%%%%%%%%%%%%%%%%%%%%%%%%%%%%%%%%%%%%%%%%%

%%%%%%%%%%%%%%%%% APPENDICES %%%%%%%%%%%%%%%%%%%%%

%\appendix



%%%%%%%%%%%%%%%%%%%%%%%%%%%%%%%%%%%%%%%%%%%%%%%%%%

% Don't change these lines
\bsp	% typesetting comment
\label{lastpage}
\end{document}